\documentclass[aps,showpacs,twocolumn,superscriptaddress,prb,floatfix]{revtex4-1}
\usepackage{graphicx,color,epsfig,rotate}
\usepackage{amssymb,amsmath,bm}

\begin{document}
%\title{Impurity-induced frustration in diluted oxides: canonical transformation and $T$-matrix study}
\title{Impurity-induced frustration: low-energy model of diluted oxides}

\author{Shiu Liu}
\affiliation{Department of Physics and Astronomy, University of California, Irvine, California 92697}

\author{A. L. Chernyshev}
\email{sasha@uci.edu}
\affiliation{Department of Physics and Astronomy, University of California, Irvine, California 92697}

\date{\today}
\begin{abstract}
We provide a detailed derivation of the low-energy model for Zn-diluted La$_2$CuO$_4$ in the limit of low doping 
together with a study of the ground-state properties of that model. We consider Zn-doped La$_2$CuO$_4$ within a 
framework of the three-band Hubbard model, which closely describes high-$T_c$ cuprates on the energy scale of the 
most relevant atomic orbitals. To obtain the low-energy effective model, we first determine hybridized electronic 
states of CuO$_4$ and ZnO$_4$ plaquets within the CuO$_2$ planes. Qualitatively, we find that the hybridization 
of zinc and oxygen orbitals can result in an impurity state with the energy $\varepsilon$, which is 
lower than the effective Hubbard gap $U$. In the limit of the effective 
hopping integral $t\ll\varepsilon$, $U$, the low-energy, spin-only Hamiltonian 
includes terms of the order  $t^2/U$ and $t^4/\varepsilon^3$. That is, besides the usual 
nearest-neighbor superexchange $J$-terms of order $t^2/U$, the low-energy model contains impurity-mediated,
further-neighbor frustrating interactions among the Cu spins surrounding Zn-sites in an otherwise unfrustrated 
antiferromagnetic background. These terms, denoted as $J'_{\rm Zn}$ and $J''_{\rm Zn}$, are of order $t^4/\varepsilon^3$ 
and can be substantial when $\varepsilon \sim U/2$, the latter value corresponding to the realistic CuO$_2$ 
parameters. In order to verify this spin-only model, we subsequently apply the $T$-matrix approach to study 
the effect of impurities on the antiferromagnetic order parameter. Previous theoretical studies, which include 
only the dilution effect of impurities, show a large 
discrepancy with experimental data in the doping dependence of the staggered magnetization at low doping. 
We demonstrate that this discrepancy is eliminated by including impurity-induced frustrations into the effective spin model 
with realistic CuO$_2$ parameters. Recent experimental study
shows a significantly stronger suppression of spin stiffness in the case of Zn-doped  
La$_2$CuO$_4$ compared to the Mg-doped case and thus gives a 
strong support to our theory. We argue that the proposed impurity-induced frustrations should be important in 
other strongly correlated oxides and charge-transfer insulators.
\end{abstract}
%--------------------------------------------------------------------
\pacs{75.10.Jm, 	% Quantized spin models
      75.40.Gb,     % Dynamic properties
      75.30.Ds,   % Spin waves
      75.50.Ee 	  % Antiferromagnetics
}
%--------------------------------------------------------------------

\maketitle
%--------------------------------------------------------------------
\section{Introduction}
%--------------------------------------------------------------------
Impurity effects in spin systems have attracted considerable attention since the 1960s
as a part of the broader interest in defects in solids\cite{Izyumov_66,Jones71,AG59,Villain_79} 
and also in the context of percolative phenomena,\cite{Breed_70,Harris_77,Stinchcombe,Cowley_77}
to which diluted magnets provide excellent experimental realizations. 
More recently, the discovery of high-$T_c$ superconductivity 
in insulating antiferromagnets doped by mobile charge carriers has lead to an extensive research effort
in various aspects of the problem, including role of disorder in quantum critical states,\cite{Vojta} 
quantum percolation,\cite{Sandvik02,Greven_MC} and others.
It has also been realized that impurities may serve as valuable local  probes into the properties of
strongly correlated systems in general, revealing important aspects of their electronic degrees of  
freedom.\cite{Vershinin,Balatsky}

%--------------------------------------------------------------------
\begin{figure}[t]
\includegraphics[width=0.99\columnwidth]{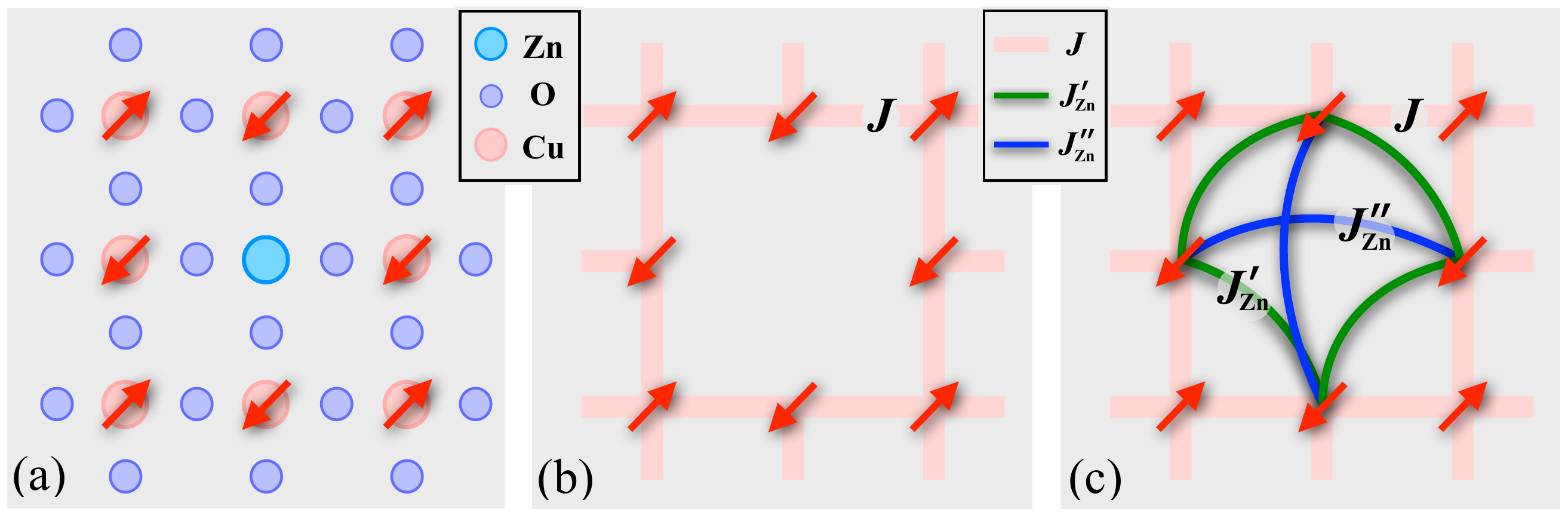}
\caption{(Color online) (a) A schematic view of Zn-doped CuO$_2$ plane. (b) Site-diluted Heisenberg model representation 
of it. The arrows are Cu spins and the lines denote superexchange interactions. 
(c) The same, with extra frustrating interactions between the next- ($J'_{\rm{Zn}}$) and the 
next-next- ($J''_{\rm{Zn}}$) nearest neighbor Cu sites, surrounding Zn impurity.}
\label{fig:JZnbonds}
\end{figure}
%--------------------------------------------------------------------

Out of the parent compounds of cuprate superconductors, it is La$_2$CuO$_4$ that has been studied
most comprehensively.\cite{cuprates,Xiao_90,Cheong_91,Corti_95,Carretta_97,Buchner_99,Greven_02,Greven_03} 
In its pristine form, it is an excellent realization of the two-dimensional, spin-$\frac12$, 
nearest-neighbor square-lattice Heisenberg antiferromagnet\cite{Manousakis_91,Dagotto_94,Chakravarty_89}  
formed by Cu$^{2+}$ ions surrounded by oxygens.
The dilution is achieved by chemically substituting isovalent Zn$^{2+}$ or Mg$^{2+}$ spinless ions for Cu$^{2+}$, 
see Fig.~\ref{fig:JZnbonds}(a). 
Thus, it is only natural to expect that the proper low-energy model of La$_2$Cu$_{1-x}$(Zn,Mg)$_x$O$_4$
must be the nearest-neighbor site-diluted Heisenberg model,\cite{Greven_02,Greven_03}  see Fig.~\ref{fig:JZnbonds}(b).
In order to elucidate the properties of La$_2$CuO$_4$ diluted by spinless impurities, and of the associated site-diluted spin 
model, extensive experimental studies have been performed using nuclear magnetic resonance (NMR) 
[nuclear quadrupole resonance (NQR)], muon spin relaxation ($\mu$SR), elastic and inelastic neutron scattering,
and magnetometry.\cite{Cheong_91,Greven_03,Greven_02,Corti_95,Carretta_97}
An equally comprehensive theoretical effort included  the spin-wave $T$-matrix, Quantum Monte Carlo, and numerical 
real-space $1/S$ calculations of a number of quantities that allowed for extensive 
cross-examinations.\cite{Bulut,kampf,Mahan,Wan,NN,Sandvik,KK_imp,Yasuda,Chernyshev_01,Chernyshev_02,Sandvik02a,Kato,Mucciolo,Delannoy_09}

One quantity in particular, the average magnetic moment per Cu, $M(x)$, has been investigated in detail.
The {\it doping dependence} of $M(x)$ is the purely quantum effect related to the impurity-induced
suppression of the order parameter because the normalization to the number of magnetic ions naturally 
separates the classical effect of dilution from it. 
While the overall results for $M$ vs $x$ 
show a reasonable agreement, a substantial discrepancy between the experiment and the theory, both analytical and QMC, 
has been observed. The experimental results indicate  a substantially stronger---a factor of approximately 2---suppression 
of the order parameter $M$ per impurity due to disorder-induced quantum fluctuations, 
and the experimental data for $M(x)$ are also always below the theoretical curves.\cite{Greven_02,Delannoy_09} 

In our recent work,\cite{Liu_09}  a resolution to this problem has been suggested:   impurities should not be considered 
as  electronically inert vacancies that simply eliminate interactions among  surrounding Cu spins. In addition to the dilution, the 
hybridized electronic states of the impurity and of the nearest oxygens can provide extra degrees of freedom that generate 
longer-range frustrating interactions, schematically shown in 
Fig.~\ref{fig:JZnbonds}(c). Such  impurity-induced frustrating interactions can be expected to significantly enhance local 
quantum fluctuations. In Ref.~\onlinecite{Liu_09} we have outlined our approach to the problem and 
presented our results for $M(x)$ together with the complementary QMC results. The latter have unequivocally  supported
the same conclusion: the {\it dilution-frustration model} exhibits stronger suppression of the order and, for a 
choice of parameters appropriate for Zn-doped La$_2$CuO$_4$,  bridges the gap between  experiments and 
previous theoretical calculations based on the {\it dilution-only model}. 
Since then  our theory has received further experimental 
confirmation from the $\mu$SR studies of   spin stiffness in Zn and Mg doped La$_2$CuO$_4$,\cite{Carretta_11} 
which demonstrated a  stronger suppression of  spin stiffness in Zn-doped case, in agreement with the expectations 
from the theory. 

In this work, we expose the details of the derivations of the dilution-frustration model and of the 
subsequent analytical calculations of the doping-dependent magnetization. 
The key element of our theory is that it starts from a realistic model, the site-diluted three-band Hubbard model, 
instead of already ``effective'' models of the CuO$_2$ plane, such as the diluted Heisenberg or the diluted one-band 
Hubbard model. 

We begin with Zn-doped La$_2$CuO$_4$ within the framework of the impurity-doped three-band Hubbard model, 
which closely describes the diluted high-$T_c$ cuprates  and other transition-metal oxides
on the energy scale of the most relevant atomic orbitals.\cite{Emery,ER} 
Similar to the derivation of the one-band Hubbard model from the three-band Hubbard 
model,\cite{ZR,Jefferson_92} the 
cell-perturbation approach is used to describe hybridization of the energy levels of Cu and Zn with oxygen orbitals. 
The approach does not require Cu-O and Zn-O hopping integrals to be smaller than the charge-transfer gap, therefore 
the locally hybridized states on Cu or Zn and surrounding O's are diagonalized without 
approximation.\cite{Jefferson_92,Chernyshev_93,Chernyshev_94_01,Chernyshev_94_02,Chernyshev_96}  
Then the three-band model can be  rewritten  as a ``multi-orbital'' Hubbard model with the effective ``Cu'' and ``Zn''
states  connected by effective hoppings. 
Since the structure of the lowest states in the single-particle and two-particle sectors 
of the multi-orbital model 
is the same as in the one-band Hubbard model ({\it i.e.} the lowest two-hole state 
is the Zhang-Rice-like singlet), the equivalence of the two models can be 
justified.\cite{Jefferson_92,Chernyshev_93,Chernyshev_94_01,Chernyshev_94_02} 
In this approach, for the dilution-only picture to be valid
the effective ``Zn''-states must not occur  below the effective Hubbard $U$, the situation  more likely
to be valid for the case of Mg-doping due to the lack of available electronic states on Mg ion. For the case of Zn-doping,
its electronic states\cite{Cheng_05,Xiang,Plakida} hybridize with the states on the oxygen orbitals and can
result in the states {\it below} the Hubbard energy gap. This provides surrounding Cu-spins with additional virtual states 
to execute their superexchange processes through, thus facilitating extra couplings that connect 
spins in the immediate vicinity of impurity.  
Therefore, the spinless impurity, in effect,  
can lead to a cage of frustrating interactions around itself, shown in Fig.~\ref{fig:JZnbonds}(c). 

For the sake of a qualitative picture, and in the spirit of mapping of the multi-band Hubbard model onto the single-band 
one, the result of this consideration is that the impurity-doped system
is {\it not} equivalent to the site-diluted Hubbard  model with electronically inert impurity sites, but
rather to the $t$-$\varepsilon$-$U$ model,  where in addition to the usual hopping $t$ and the two-particle
energy gap $U$ there is the lowest energy state of the effective impurity sector, denoted as $\varepsilon$. 
At half-filling, the $t$-$U$ part reduces to the Heisenberg model with $J\!\sim\!t^2/U$, as usual, with
the higher-order terms $\sim\! t^4/U^3$ negligible if $t\!\ll\! U$.\cite{Takahashi_77,Yoshioka_88,Chernyshev_04}
Virtual transitions through the effective impurity level $\varepsilon$ generate superexchange interactions 
of the order of $t^4/\varepsilon^3$ between the next-  and the next-next-nearest-neighbor Cu-spins
($J'_{\rm Zn}$ and $J''_{\rm Zn}$) that are also nearest-neighbors of the impurity site. 
When the energy at  the impurity site $\varepsilon$ is less than the Hubbard gap, such terms 
are not negligible and may be  comparable to $J$. For an estimate, taking $\varepsilon=U/2$  
and $U/t=10$ gives $J^\prime_{\rm Zn}/J\!\sim\! (t/U)^2(U/\varepsilon)^3\!\sim\! 0.1$, and, given
the geometry of the square lattice,  the combined effect 
per impurity is $J^{tot}_{\rm Zn}\!=\!4J^\prime_{\rm Zn}\!+\!2J^{\prime\prime}_{\rm Zn}\!\sim\! 0.6 J$.
The origin of $J'_{\rm Zn}$ and $J''_{\rm Zn}$ is clearly distinct from the generally considered next- 
and  next-next-nearest-neighbor superexchange interaction ($J_2$ and $J_3$), which are of the order of 
$t^4/U^3$ and do not affect the order parameter specifically due to dilution.\cite{Tremblay_09}

While the technical details and the effort of the present work are more involved, this qualitative $t$-$\varepsilon$-$U$ model
properly reflects the key idea of our approach. In practice, we perform a similar type of 
the 4$^{th}$-order expansion\cite{Chernyshev_04} of the multi-orbital Hubbard model, keeping track of all the 
relevant one- and two-hole states of the model and their dependencies on the original three-band model parameters. 
This allows us to perform a detailed microscopic calculations of $J^\prime_{\rm Zn}$ and
$J^{\prime\prime}_{\rm Zn}$ and estimate their values. Since the experimental value of the nearest-neighbor 
superexchange for La$_2$CuO$_4$ is known ($J\!\simeq\!0.13$eV), it can be used to narrow down the
range of parameters of the three-band model as was done previously.\cite{Chernyshev_96} 
Although the electronic parameters of Zn states, such as the  energy of the bare  Zn-level  and the Zn-O hybridization, 
are not known precisely,\cite{Plakida,Cheng_05} we vary them to verify that the  energy of the lowest impurity level 
$\varepsilon$ indeed falls comfortably below the effective Hubbard $U$ for a wide and reasonable range of both
parameters. By projecting the multi-orbital Hubbard model onto the low-energy spin-only model, we analyze  
possible value of the total frustrating effect per impurity
and find it to be between $J^{tot}_{\rm Zn}\sim (0.2-1.0) J$ for the same range of parameters.

In addition, we also provide a detailed analysis of  the individual processes that 
contribute to $J'_{\rm Zn}$ and $J''_{\rm Zn}$. Specifically, we have found that, counterintuitively,  the longer-ranged $J''_{\rm Zn}$ 
is greater than $J'_{\rm Zn}$.
This is due to a subtle cancellation  between the ``regular'' 4$^{th}$-order superexchange processes and
the analogs of the ring-exchange-type processes involving Zn and three Cu sites on a nearest-neighbor  
plaquette, see Fig.~\ref{fig:JZnbonds}(a), with the latter contributing to $J'_{\rm Zn}$, but not to $J''_{\rm Zn}$.
That results in a {\it stronger} frustrating coupling between copper spins
across the Zn-site, with the ratio  
$J^{\prime\prime}_{\rm Zn}/J^{\prime}_{\rm Zn}\simeq 2 \sim 4$ in a wide range
of the three-band model parameters. 
Altogether, the calculation based on the three-band model, albeit more involved, provides a strong support to
our central idea and gives an order-of-magnitude estimate of the frustrating terms in the dilution-frustration model.

Upon establishing the structure of the low-energy spin-only model for the diluted system, to which we refer
to as to the {\it dilution-frustration model},  we investigate the impact of impurities in this model on
the order parameter $M$. This is achieved  by means of the analytical $T$-matrix approach within 
the spin-wave approximation based on the exact diagrammatic treatment of the impurity scattering amplitudes 
and subsequent disorder averaging. Technically, we closely follow the 
approach of Refs.~\onlinecite{Chernyshev_01,Chernyshev_02}.

First, we apply  the spin-wave approximation to rewrite the dilution-frustration model on the 2D square lattice. 
After the Fourier and Bogolyubov transformation we decompose the impurity 
scattering matrices into the $s$-, $p$-, $d$-wave
orthogonal components with respect to the scattering site. 
Then, the $T$-matrix approach is used to solve exactly  the problem of scattering off one impurity in each
of the scattering channels. The subsequent disorder-averaging  approximates impurities as independent random 
scatterers and effectively restores translational invariance for spin excitations propagating
in an effective medium. Such an approximation neglects impurity-impurity interaction effects, which are expected to be small at
small doping. Disorder averaging extends the $T$-matrix approach to the finite impurity concentration and yields 
the spin-wave self-energies as simply related to the forward scattering components of the $T$-matrix. 
Next, for the given impurity concentration and values of frustrating parameters, the on-site ordered magnetic moment $M$ is calculated from the renormalized magnon  Green's functions.

Compared to the dilution-only model, the modification of this method for the dilution-frustration model  
concerns changes in the $p$- and $d$-wave scattering channels, while the $s$-wave channel can be shown 
to be unaffected by the frustrating terms. 
Generally, the advantage of the $T$-matrix method is that it offers a systematic way of studying the order parameter 
$M$ as a function of the concentration $x$ and of the parameters $J^\prime_{\rm Zn}$ 
and $J^{\prime\prime}_{\rm Zn}$.

One of the unusual findings in this study is that the next-next-nearest neighbor $J^{\prime\prime}_{\rm Zn}$ frustrating bond 
suppresses the order as effectively as two next-nearest $J^{\prime}_{\rm Zn}$ bonds 
of the same strength. This result has also been supported by the QMC calculations, as was discussed previously.\cite{Liu_09} 
Given that according to the three-band model calculations the $J^{\prime\prime}_{\rm Zn}$-term is larger than the 
$J^{\prime}_{\rm Zn}$-term, this finding underscores its importance for the mechanism of impurity-enhanced suppression
of the order parameter.

We find that the experimental rate of suppression of the order 
in Zn-doped La$_2$CuO$_4$ is met by the $T$-matrix results 
of the dilution-frustration model  at $J_{\rm Zn}^{tot}\!=\!0.28J$
if we fix the ratio of the two frustrating terms to 2, $J^{\prime\prime}_{\rm Zn}\!=\!2J^{\prime}_{\rm Zn}\!=\!0.07J$, 
or at $J_{\rm Zn}^{tot}\!=\!0.24J$ for $J^{\prime\prime}_{\rm Zn}\!=\!4J^{\prime}_{\rm Zn}\!=\!0.08J$,
according to the three-band model  results.  As was discussed in our previous work,\cite{Liu_09}
QMC results seem to suggest a somewhat higher value $J_{\rm Zn}^{tot}\!\agt\!0.4J$ 
($J^{\prime\prime}_{\rm Zn}\!=\!2J^{\prime}_{\rm Zn}\!\agt\!0.1J$), although they are obtained from the finite-size 
extrapolations that may overestimate $J_{\rm Zn}^{tot}$. Both the $T$-matrix and the QMC results correspond to  
a modest amount of frustration, well within the window suggested by the  three-band model calculations.
With our  analytical and numerical  results  agreeing quantitatively  with each other, we have suggested 
further high-precision experiments at low doping. 
One of such  experimental confirmations has come recently from the $\mu$SR studies of Zn- and 
Mg-doped La$_2$CuO$_4$,\cite{Carretta_11} 
which has shown a substantially stronger suppression of the  spin stiffness in the case of Zn-doping, 
in agreement with the expectations from our theory. 

The paper is organized as follows. In Sec.~\ref{Sec_II} we discuss the diluted 
three-band Hubbard model within the cell-perturbation approach and 
derive the multi-orbital Hubbard model from it. We study the hybridized impurity level  and analyze its
energy with respect to the effective Hubbard gap for a range of three-band model parameters. 
We derive an effective low-energy  spin-only model by applying canonical transformation 
to the multi-orbital Hubbard model at half-filling. The virtual steps leading to the frustrating 
superexchange interactions, $J'_{\rm Zn}$ and $J''_{\rm Zn}$, are analyzed and their values are calculated.  
In Sec.~III, we study the effective 
dilution-frustration model  using the $T$-matrix approach and  disorder averaging. We derive the 
staggered magnetization as a function of impurity concentration and frustrating couplings. The results are compared 
with experimental data.  Section IV contains our conclusions. Appendices A and B contain details of the 
three-band model and the $T$-matrix calculations, respectively.

%--------------------------------------------------------------------
\section{effective model of copper-oxide plane with zinc impurities}
\label{Sec_II}
%--------------------------------------------------------------------

In this section, we derive an effective low-energy model for the Zn-doped La$_2$CuO$_4$ at half-filling. 
We begin with the consideration of the realistic, Zn-diluted three-band Hubbard model, which contains additional 
impurity states associated with Zn. Using Wannier-orthogonalization for O-orbitals and 
a more natural language of the locally hybridized CuO$_4$ and ZnO$_4$ states, we 
first rewrite the three-band Hamiltonian as the multi-orbital Hubbard model. 
We discuss the new feature of the model---the hybridized Zn and O orbitals 
can result in an impurity state with the energy that is lower than the effective Hubbard gap. 
The subsequent transformation to the low-energy, spin-only model is known as the cell-perturbation method. 
The local basis of CuO$_4$ and ZnO$_4$ states of the multi-orbital Hamiltonian provides a  natural small parameter for 
such a projection, the effective hopping between the states of the nearest-neighbor clusters.
We proceed with this transformation and give a quantitative analysis of  the individual processes that 
contribute to the superexchange  terms of the effective model. Next, we analyze possible 
range of the frustrating interactions in the  low-energy model of the Zn-doped CuO$_2$ plane.
The discussed approach should be valid as long as the system is 
in the Mott insulating state.

%--------------------------------------------------------------------
\subsection{Three-band to multi-orbital Hubbard model}
%--------------------------------------------------------------------

The three-band Hubbard model, which was proposed to describe relevant electronic degrees of freedom of the 
CuO$_2$ planes of the high-$T_c$ cuprates,\cite{Emery} is formulated in terms of the hole creation and annihilation
operators acting on the vacuum state of the completely filled $3d^{10}$ shells of Cu and $2p^6$ shells of O 
\begin{eqnarray} 
&&{\cal H} = \varepsilon_d \sum_{l\alpha} n^d_{l\alpha}+\varepsilon_p \sum_{m\alpha} n^p_{m\alpha}
+U_d \sum_l n^d_{l\uparrow}n^d_{l\downarrow}\nonumber\\
&& \phantom{{\cal H} = } -t_{pd}\sum_{\langle lm\rangle\alpha}
\left( d^\dagger_{l\alpha}p^{\phantom\dag}_{m\alpha}+ {\rm H.c.}\right)\, ,
\label{H:three-band-Cu}
\end{eqnarray}
where $\varepsilon_d$ and $\varepsilon_p$ are the energies of the hole in the copper $d_{x^2-y^2}$ and in the oxygen 
$p_x$ or $p_y$ orbitals, respectively. The copper(oxygen) sites are labeled with the index $l$($m$), the number 
operators are $n^d_{l\alpha}=d^\dagger_{l\alpha}d^{\phantom\dag}_{l\alpha}$
($n^p_{m\alpha}=p^\dagger_{m\alpha}p^{\phantom\dag}_{m\alpha}$), and $\alpha=\uparrow,\downarrow$ is the spin. 
The Hubbard repulsion 
in the copper $d_{x^2-y^2}$ orbitals is $U_d$ and the hopping between copper and oxygen orbitals is $t_{pd}$. 
For the Cu-O hoppings we use the convention\cite{ER} in which 
the relative signs of orbitals are absorbed into the definition of $p_{m\alpha}^{\phantom\dag}$ 
($p_{m\alpha}^{\dag}$) operators.
The summation $\langle lm\rangle$ is over the nearest-neighbor Cu-O bonds. 

\begin{figure}[t]
\includegraphics[width=0.7\columnwidth]{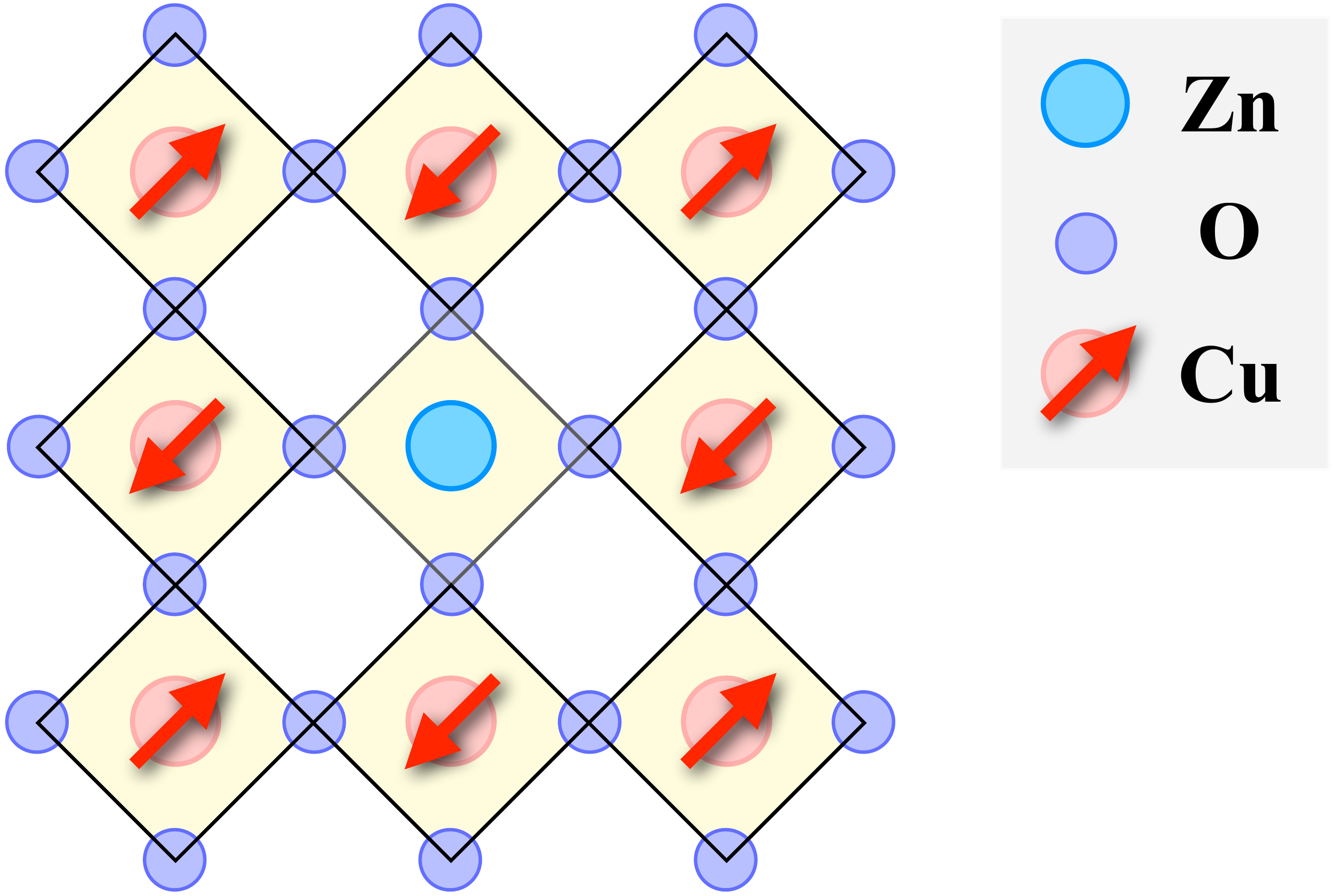}
\caption{(Color online) Zn-doped CuO$_2$ plane with natural partitioning in CuO$_4$ and ZnO$_4$ clusters. }
\label{fig:cluster}
\end{figure}

The relevant transitions within the three-band model involve   $d^{10}$-$d^8$ states on  copper 
and  $p$ states on  oxygen sites.  
At half-filling, there is one hole per CuO$_2$ unit cell and the ground state is the antiferromagnetic
insulator.\cite{Dagotto_94} The localized $S=1/2$ spins, forming the 
N\'{e}el-ordered state on the square lattice, are provided by the  holes that are predominantly 
in the $d^9$ states on Cu and are hybridized with the $p^5$ states on O, see Fig.~\ref{fig:cluster}.

The minimal form of the three-band model in (\ref{H:three-band-Cu}) is often supplemented with   additional terms, 
such as the direct oxygen-oxygen hopping, on-site Coulomb interactions on O-sites, and the nearest-neighbor repulsion 
between O- and Cu-holes. 
The main effect of such terms is in a quantitative renormalization of the energy levels and  hopping 
integrals,\cite{Jefferson_92,Chernyshev_94_01} leaving the results obtained with (\ref{H:three-band-Cu}) qualitatively 
the same. 

%-------------------------------------------------
%-------------------------------------------------

The substitution of the isovalent Zn$^{2+}$  ion  with the nominally completely filled $3d$-shell for Cu$^{2+}$
 leaves the oxygen lattice  translationally invariant, see Fig.~\ref{fig:cluster}. 
While the electronic levels of Zn doped into a CuO$_2$ plane have not been determined precisely,\cite{Cheng_05,Xiang,Plakida}  
it is generally agreed that the relevant states may occur in a reasonable vicinity of the oxygen level,
$\varepsilon_{\rm Zn}-\varepsilon_{p}\simeq 2\!-\!5$eV, and we treat this energy as an adjustable parameter in the
following. Thus, the impurity Hamiltonian can be written as
\begin{eqnarray}
&&\delta {\cal H}_{\rm Zn} = \varepsilon_{\rm Zn}\sum_{\ell\alpha} n^{\rm Zn}_{\ell\alpha}+
U_{\rm Zn}\sum_\ell n^{\rm Zn}_{\ell\uparrow}n^{\rm Zn}_{\ell\downarrow}\nonumber\\
&&\phantom{\delta {\cal H}_{\rm Zn} =} -t_{\rm ZnO} \sum_{\langle \ell m\rangle\alpha}
\left(a^\dagger_{\ell\alpha}p^{\phantom\dag}_{m\alpha}
+ {\rm H.c.}\right)\, ,
\label{H:three-band-Zn}
\end{eqnarray}
where  Zn sites are labeled with $\ell$, $\varepsilon_{\rm Zn}$ is the energy
of the relevant orbital on Zn,  $U_{\rm Zn}$ is an effective Hubbard repulsion in that orbital, 
and $t_{\rm ZnO}$ is the hopping integral 
between nearest-neighbor Zn and O sites, yet another adjustable parameter.  The hole creation/annihilation operators on 
Zn are $a^\dagger_{\ell\alpha}$ and $a^{\phantom\dag}_{\ell\alpha}$ 
and $n^{\rm Zn}_{\ell\alpha}=a^\dagger_{\ell\alpha}a^{\phantom\dag}_{\ell\alpha}$. 
	
The checkerboard structural motif of the  CuO$_2$ plane in Fig.~\ref{fig:cluster}  suggests 
a natural partitioning of the localized states 
into symmetric CuO$_4$ and ZnO$_4$ clusters, in which each  Cu or Zn is surrounded by four 
oxygens.\cite{ZR} The advantage of such basic units is in constructing the local states that allow to take into account 
local hybridization of Cu or Zn and surrounding O states without approximation, while the remaining hybridization between 
the clusters can be treated perturbatively. In particular, such a cell-perturbation approach does 
not require Cu-O and Zn-O hopping integrals to be much smaller than the charge-transfer gap $\Delta=\varepsilon_{p}-\varepsilon_{d}$.\cite{Jefferson_92,Chernyshev_93,Chernyshev_94_01,Chernyshev_94_02,Chernyshev_96}

On the other hand, the linear combinations of the oxygen states in CuO$_4$ (ZnO$_4$) cluster are not orthogonal to the ones
in the nearest-neighbor clusters. The elegant Wannier-orthogonalization procedure, suggested in Ref.~\onlinecite{ZR},  
treats  the O-lattice separate from the Cu and Zn, and, via orthogonalizing $p$-operators in the ${\bf k}$-space,
leads to the basis of the symmetric-oxygen orbitals  
\begin{eqnarray}
\big\lbrace p_{m,\alpha}^x ,\; p_{m\alpha}^y \big\rbrace\rightarrow q_{l,\alpha}(q_{\ell,\alpha})\, ,
\label{Wannier}
\end{eqnarray}
that are now associated with the same site index as the Cu (Zn) of the CuO$_4$ (ZnO$_4$) cluster. 
The details of this procedure are given in Appendix~\ref{AppA}. 

With that, it is natural to divide the Cu-O (Zn-O) hopping terms in (\ref{H:three-band-Cu}) and (\ref{H:three-band-Zn}) into 
the local part, which corresponds to hybridization within one cluster,  and into the 
hopping part, corresponding to the coupling between the states in different clusters.
We note, that due to the non-local nature of the  Wannier-orthogonalization, the hopping parts now contain terms
that go beyond the nearest-neighbor clusters. Thus, the Hamiltonian of (\ref{H:three-band-Cu}), written using the orthogonalized 
basis of O-states (\ref{Wannier}) is
\begin{eqnarray}
&&{\cal H}^{{\rm loc}} = \sum_{l\alpha}\bigg\lbrace\varepsilon_d n^d_{l\alpha} + \varepsilon_p n^q_{l\alpha}
+\frac{U_d}{2} n^d_{l\alpha}n^d_{l\bar{\alpha}}\nonumber\\
&&\phantom{{\cal H}^{{\rm loc}} = \sum_{l\alpha}\bigg\lbrace}
-2\lambda_0 t_{pd}\left( d^\dagger_{l\alpha}q^{\phantom\dag}_{l\alpha}+ {\rm H.c.}\right) \bigg\rbrace 
\label{H:symmetric_oxygen_Cu}\\
&&{\cal H}^{{\rm hop}} =  -\sum_{\langle ll'\rangle\alpha} 2\lambda_{ll'}t_{pd}
\left( d^\dagger_{l\alpha}q^{\phantom\dag}_{l'\alpha}+ {\rm H.c.}\right) \, , \nonumber
\end{eqnarray}
where $\lambda_{ll'}$ are the Wannier amplitudes given in (\ref{eq:lambda_ll'}). 
First, the amplitudes $\lambda _{ll'}$ decrease rapidly with distance,\cite{Chernyshev_93}   
$\lambda _{ll'}\sim 1/|{\bf r}_l - {\bf r}_{l'}|^3$, so the terms beyond the nearest neighbor in the hopping part of 
(\ref{H:symmetric_oxygen_Cu}) can be neglected. Second,  the intra-cluster amplitude $\lambda_0 = 0.9581$ 
is close to 1, and thus is taking most of the hybridization into account. The nearest-neighbor amplitude
$\lambda_{\langle ll'\rangle} =0.1401$ leads to the effective  hopping parameter between 
different states that is significantly reduced compared to the bare $t_{pd}$ hopping. This 
provides a strong justification for the subsequent perturbative expansion in such an effective hopping relative to
the effective Hubbard gap, the latter largely defined by the local hybridization. 

Applying the same transformation of the O-states (\ref{Wannier}) to the Hamiltonian of the Zn-cluster 
(\ref{H:three-band-Zn}) we obtain
\begin{eqnarray}
&&\delta {\cal H}_{\rm Zn}^{{\rm loc}} =\sum_{\ell\alpha}\bigg\lbrace 
\varepsilon_{\rm Zn} n^{\rm Zn}_{\ell\alpha}+\varepsilon_p n^q_{\ell\alpha}
+\frac{U_{\rm Zn}}{2} n^{\rm Zn}_{\ell\alpha}n^{\rm Zn}_{\ell\bar{\alpha}}
 \nonumber\\
&&\phantom{\delta {\cal H}_{\rm Zn}^{{\rm loc}} =\sum_{\ell\alpha}\bigg\lbrace}
-2\lambda_0 t_{\rm ZnO}\left(a^\dagger_{\ell\alpha}q^{\phantom\dag}_{\ell\alpha}+ {\rm H.c.}\right)  
\bigg\rbrace
\label{H:symmetric_oxygen_Zn}\\
&&\delta {\cal H}_{\rm Zn}^{{\rm hop}} = -\sum_{\langle\ell l\rangle\alpha} 2\lambda_{\ell l}t_{\rm ZnO}
\left( a^\dagger_{\ell\alpha}q^{\phantom\dag}_{l\alpha}+ {\rm H.c.}\right)\, .  \nonumber
\end{eqnarray}
Note that even if Zn-states are completely neglected  in 
$\delta {\cal H}_{\rm Zn}^{{\rm loc}}$, ZnO$_4$ cluster still has an unoccupied oxygen state\cite{Plakida} 
with the energy $\varepsilon_p$, which is higher than the hybridized CuO$_4$ states of the 
surrounding clusters, and  it permits virtual superexchange processes through it. 

%----------------------------------------------------
\begin{figure}[t]
\includegraphics[width=0.9\columnwidth]{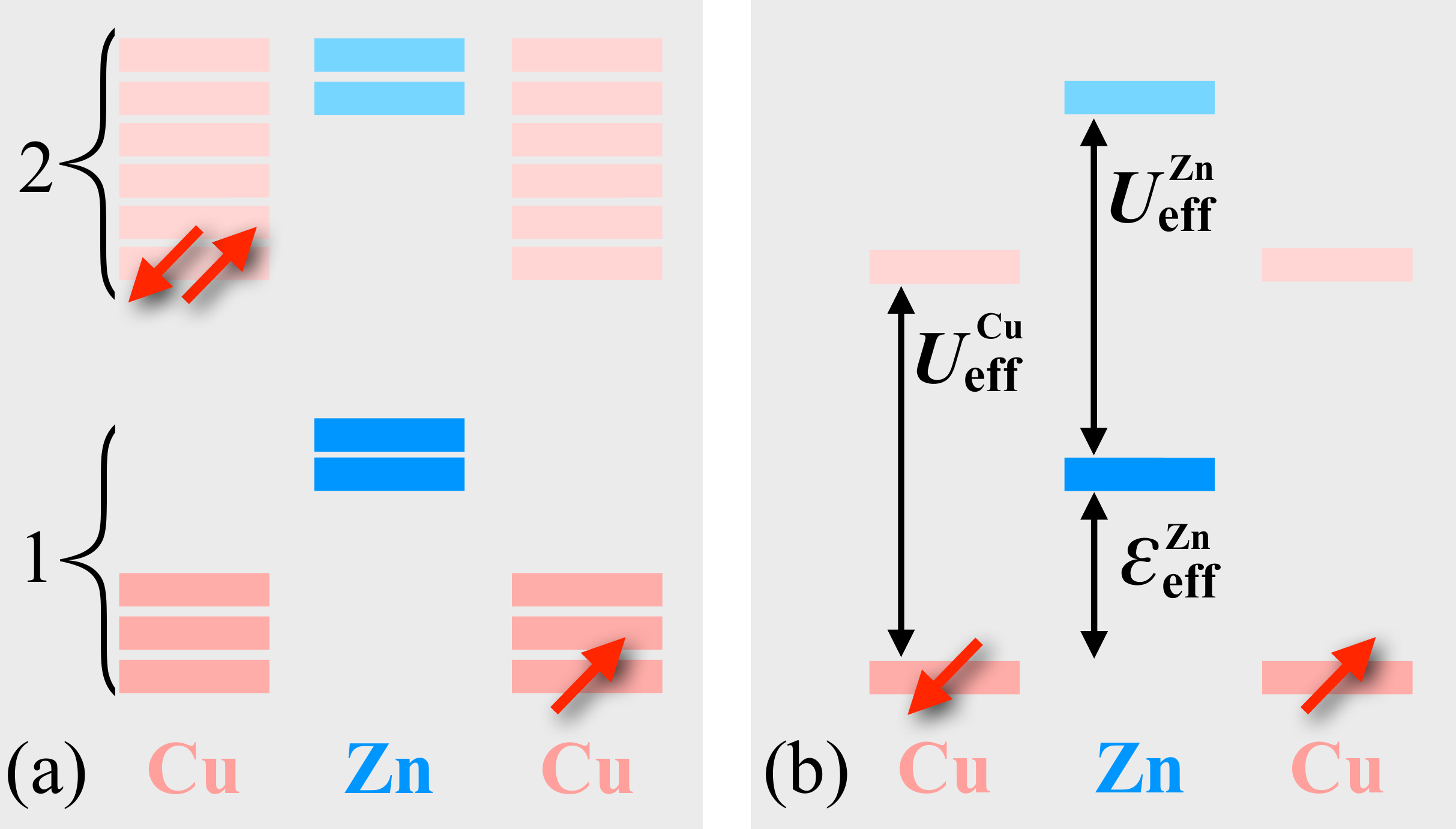}
\caption{(Color online) (a) A schematic view of the energy levels 
$E_{i}$ and $\tilde{E}_i$ of one- and two-particle sectors in Cu and Zn clusters in Eqs. (\ref{H:diagonalized_Cu}) and 
(\ref{H:diagonalized_Zn}). 
(b) The levels of the effective $t-\varepsilon-U$ Hamiltonian that keeps  the lowest energy 
states in each  sector.}
\label{fig:energy_level}
\end{figure}
%----------------------------------------------------

The next step in the cell-perturbation approach is the diagonalization of 
${\cal H}^{{\rm loc}}$ and $\delta{\cal H}^{{\rm loc}}$.
Such a diagonalization is performed separately for the states with different number of holes in a cluster. 
Because the system is close to the half-filling, the most relevant states are the one- and two-hole states, 
and the states with more holes are higher in energy, see Fig.~\ref{fig:energy_level}(a). 
The full set of  $d_l$, $a_\ell$, and $q_l$ one- and two-hole states of 
Cu- and Zn-clusters  are listed in Appendix~\ref{AppA}, 
 (\ref{eq:Cu_one_hole_states})-(\ref{eq:Zn_two_hole_triplets}), see also Ref.~\onlinecite{Chernyshev_94_01}.

After the diagonalization of the local parts of the Hamiltonian, the hopping terms are also rewritten in the basis of 
the new states. Altogether, the three-band model ${\cal H}$ of (\ref{H:symmetric_oxygen_Cu}) can be rewritten as 
a ``multi-orbital'' Hubbard model with the effective ``Cu'' 
eigenstates  with local energies $E_i$ connected by effective hoppings $F_{ii'}^{jj'}$:
\begin{eqnarray}
&&{\cal H} = \sum_{li} E_{i} |\psi_l^{i}\rangle \langle \psi_l^{i}|
\label{H:diagonalized_Cu}\\
&&\phantom{{\cal H}}+\sum_{\langle ll'\rangle}\sum_{ii'jj'} F_{ii'}^{jj'} 
\left( |\psi_l^{i}\rangle|\psi_{l'}^{i'}\rangle\langle\psi_{l'}^{j'}|\langle\psi_{l}^{j}| + {\rm H.c.}\right)\, .\nonumber
\end{eqnarray}
Similarly, for $\delta{\cal H}$ of (\ref{H:symmetric_oxygen_Zn})
\begin{eqnarray}
&&\delta {\cal H} = \sum_{\ell i} \tilde{E}_i |\tilde{\psi}_\ell^i\rangle \langle \tilde{\psi}_\ell^i|
\label{H:diagonalized_Zn}\\
&&\phantom{\delta {\cal H}}+\sum_{\langle \ell l\rangle}\sum_{ii'jj'}\tilde{F}^{jj'}_{ii'} \left( |\tilde{\psi}_\ell^i\rangle|\psi_{l}^{i'}\rangle\langle\psi_{l}^{j'}|\langle\tilde{\psi}_{\ell}^{j}| + {\rm H.c.}\right)\,.\nonumber
\end{eqnarray}
The hybridized, orthogonal sets of local states are $|\psi_l^i\rangle$ for CuO$_4$ and 
$ |\tilde{\psi}_\ell^i\rangle$  for ZnO$_4$ clusters, where $l(\ell)$ is the site-index and $i$ is 
labeling the states. 
For example, the three one-hole states in Cu cluster in Fig.~\ref{fig:energy_level}(a) are the 
``bonding'' and ``antibonding'' mix of Cu and symmetric O states $d_l$ and $q_l$, 
Eqs.~(\ref{eq:qs}) and (\ref{eq:Cu_one_hole_states}), and the antisymmetric O-state ($\tilde{q}_l$, Eq. (\ref{eq:qs})). 
The hopping integrals between adjacent clusters, $F_{ii'}^{jj'}$ and $\tilde{F}^{jj'}_{ii'}$, connect
initial and final states $i$, $i'$ and $j$, $j'$. 

The Hamiltonian in (\ref{H:diagonalized_Cu}) and (\ref{H:diagonalized_Zn}) is ``multi-orbital'' because each $n$-particle sector
contains more than one state. However, since such states are locally orthogonal and we are interested 
in the low-energy effective model of (\ref{H:diagonalized_Cu}) and (\ref{H:diagonalized_Zn}), the most 
important transitions involve the {\it lowest} states from each sector, see Fig.~\ref{fig:energy_level}(a). 
Importantly, the structure of the lowest states in the single-particle and two-particle sectors 
of the multi-orbital model is the same as in the one-band Hubbard model, justifying the close correspondence of 
the  three-band and one-band Hubbard 
models.\cite{Jefferson_92,Chernyshev_93,Chernyshev_94_01,Chernyshev_94_02}  
The lowest one-hole states in Cu- and Zn-cluster are $S=1/2$ doublets, the linear combinations of the oxygen and the 
Cu(Zn)-orbitals, while the lowest two-hole state 
is the Zhang-Rice-like singlet, a mix of three different singlets [Cu-Cu, O-O and Cu-O], see 
 (\ref{eq:Cu_one_hole_states})-(\ref{eq:Zn_two_hole_singlets}).

The new physics is brought in by the possibility of an unoccupied impurity level, 
the lowest state from the ``Zn'' single-particle sector, 
to be lower than the effective Hubbard $U$ on the ``Cu'' sites.  This leads to an effective $t-\varepsilon-U$ model, see
Fig.~\ref{fig:energy_level}(b), which provides surrounding Cu-spins with additional virtual channel for the  
superexchange processes, connecting  spins in the vicinity of  impurity.  
Conversely, if the effective ``Zn''-states do not occur  below the Hubbard gap (the case of 
Mg$^{2+}$ due to the lack of available states on it), the impurity can be treated as electronically inert, leading 
only to dilution, but not to extra interactions. 
Parameters that control the impurity level are discussed next.

%-------------------------------------------------------------------------
\subsection{Parameters and effective impurity level}
%-------------------------------------------------------------------------

The realistic values of the three-band model parameters for the CuO$_2$ plane
have a certain degree of variation because none of them is measurable directly and they 
come as a result of a parametrization of the first-principles calculations.
Thus, the Hubbard repulsion on Cu is $U_d\!=\!8\!-\!12$eV, the charge-transfer gap 
$\Delta\!=\!\varepsilon_p \!-\! \varepsilon_d\!=\!2\!-\!4$eV,
and the Cu-O hopping is $t_{pd}\!=\!1\!-\!1.5$eV.\cite{Jefferson_92,Chernyshev_94_01,Chernyshev_94_02}  
A physical approach to fix them to a particular set of values was suggested in Ref.~\onlinecite{Chernyshev_94_02}.
One can require that the observables, such as the Cu-Cu superexchange or the optical gap, calculated from the model
(\ref{H:diagonalized_Cu}) match their observed values.  
In our case, we fix them to $U_d(U_{\rm Zn})\!\gg\!\Delta$, $\Delta \!=\! 3$eV, and $t_{pd} \!=\! 1.5$eV to yield  
the experimental value of the superexchange $J\!\simeq\!0.13$eV in the effective low-energy model  
discussued in Sec.~\ref{Sec_II}.C.  
Among other things, the effective Hubbard gap for the Zhang-Rice-like singlet and 
the ground state half-filled system is $U_{\rm eff}\!\approx\! 3.6$eV.

\begin{figure}[t]
\includegraphics[width=1.0\columnwidth]{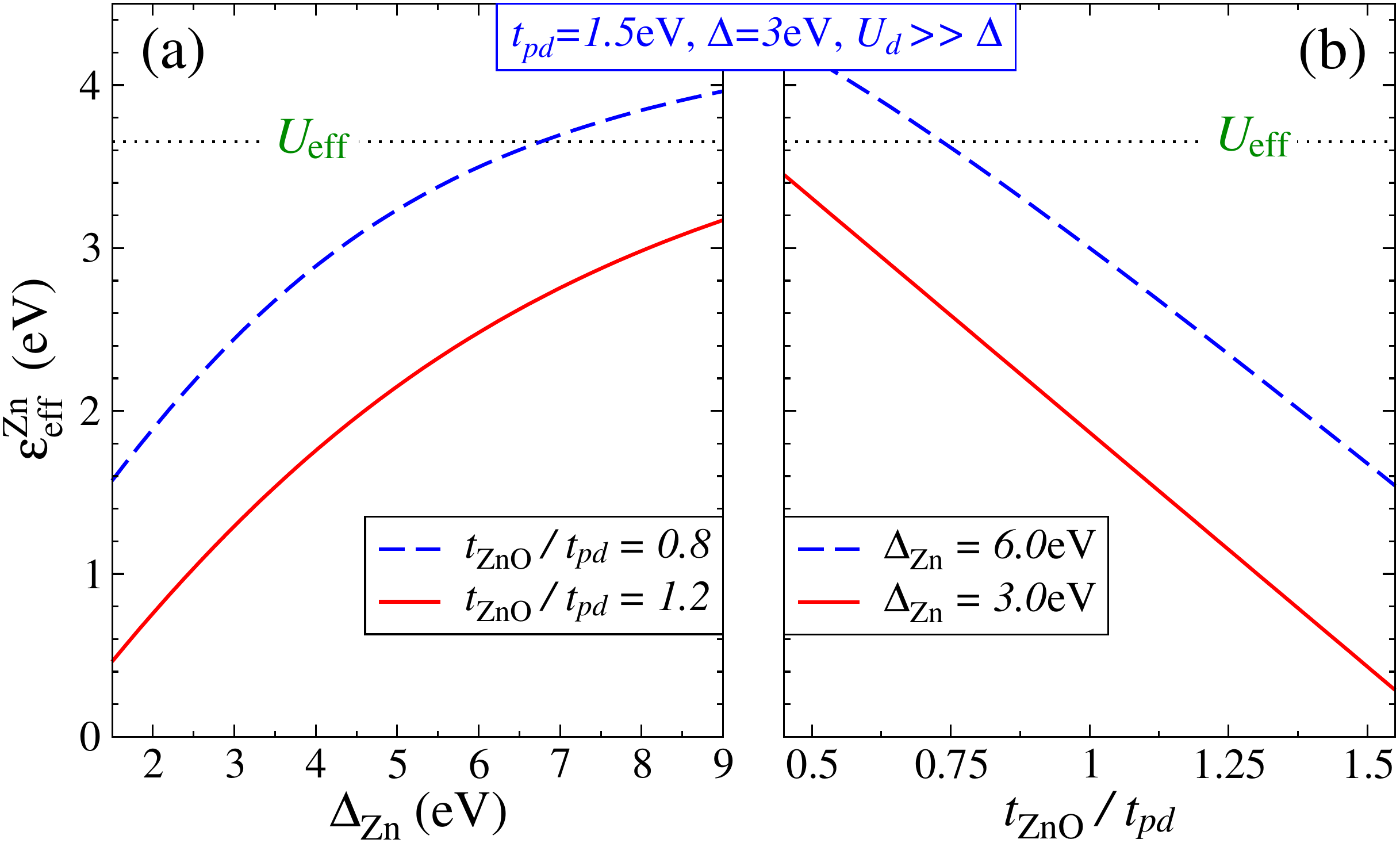}
\caption{(Color online) (a) Energy of the lowest state from the Zn-cluster single-particle sector, $\varepsilon_{\rm eff}^{\rm Zn}$, 
vs the energy difference between the bare one-hole levels on Zn and O, 
$\Delta_{\rm Zn}\!=\!\varepsilon_{\rm Zn}\!-\!\varepsilon_p$, for 
two values of $t_{\rm ZnO}$. (b) $\varepsilon_{\rm eff}^{\rm Zn}$ vs $t_{\rm ZnO}/t_{pd}$ for two representative values of 
$\Delta_{\rm Zn}$.
Parameters of the three-band model are fixed as described in text, $t_{pd}\!=\!1.5$eV, $\Delta\!=\!3$eV, $U_d\!\gg\!\Delta$.
 The resultant Hubbard gap is $U_{\rm eff}\approx 3.6$eV.}
\label{fig:eff_epsilon}
\end{figure}

The parameters of Zn states, such as the bare one-hole level on Zn, 
$\Delta_{\rm Zn}\!=\!\varepsilon_{\rm Zn}\!-\!\varepsilon_p$, and the Zn-O hopping $t_{\rm ZnO}$,
are not precisely known.\cite{Cheng_05,Xiang,Plakida}
As is shown in Fig.~\ref{fig:eff_epsilon}, 
we vary them substantially to determine the resulting energy of the lowest level in the single-particle sector of the 
ZnO$_4$ cluster, $\varepsilon_{\rm eff}^{\rm Zn}$.
In Fig.~\ref{fig:eff_epsilon}(a) we show its dependence on the  energy of the bare 
Zn-level, $\Delta_{\rm Zn}$, for two representative values of $t_{\rm ZnO}$,
and in Fig.~\ref{fig:eff_epsilon}(b) on the the hybridization $t_{\rm ZnO}$ for two representative values of $\Delta_{\rm Zn}$.
We also show the effective Hubbard energy $U_{\rm eff}$ to demonstrate the validity of the qualitative
level structure in Fig.~\ref{fig:energy_level} for a wide range of parameters. Altogether, 
the electronic levels of Zn and their hybridization with O-states are important  
in lowering $\varepsilon_{\rm eff}^{\rm Zn}$, and $\varepsilon_{\rm eff}^{\rm Zn}$ is below the Hubbard gap for 
the realistic parameters of the model. 

%--------------------------------------------------------------
%--------------------------------------------------------------

%--------------------------------------------------------------
\begin{figure}[b]
\includegraphics[width=0.65\columnwidth]{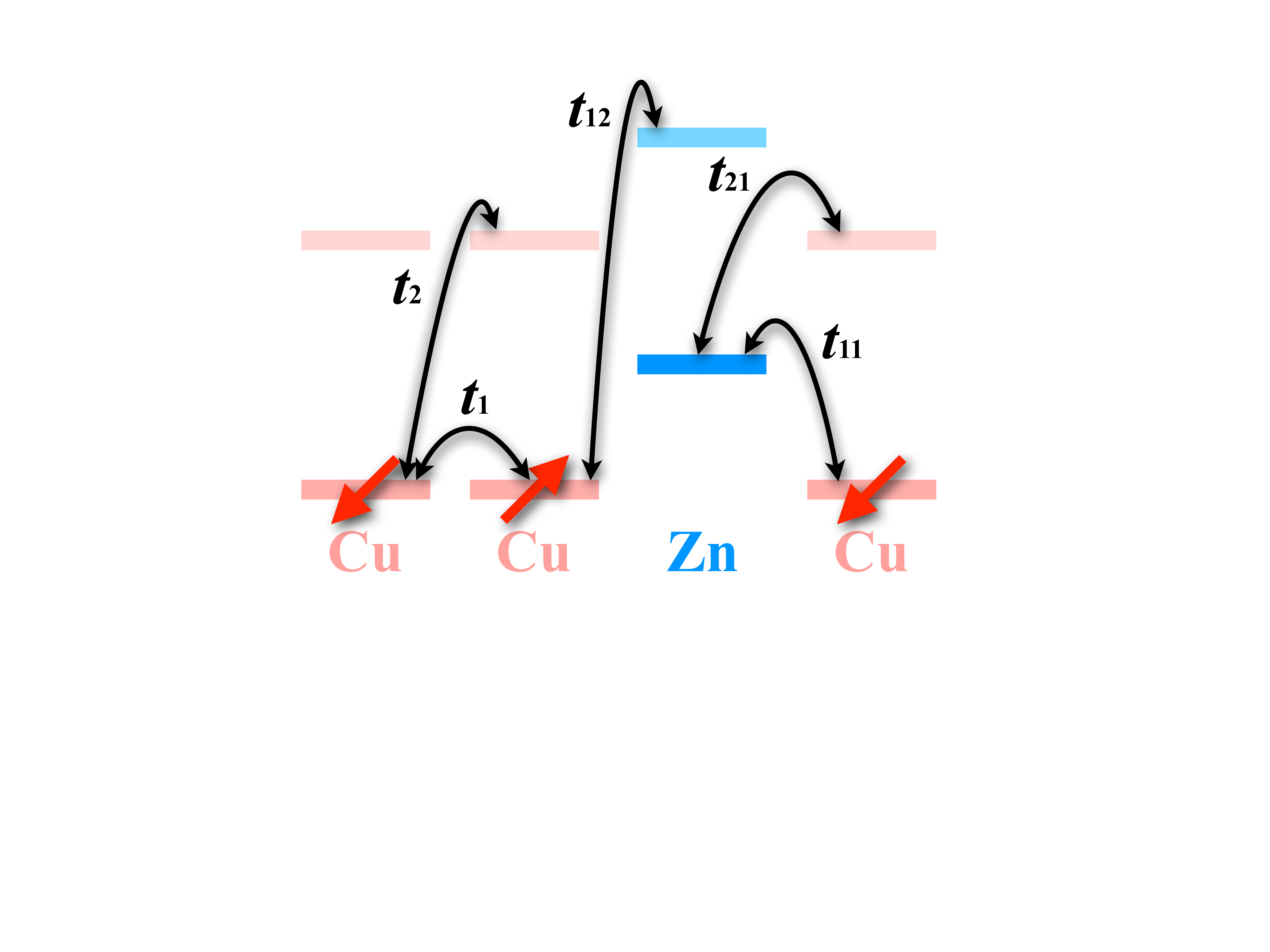}
\caption{(Color online) Schematic view of various virtual hopping processes between relevant energy levels 
of  the model (\ref{H:diagonalized_Cu}) and (\ref{H:diagonalized_Zn}), see text for notations.}
\label{fig:effHop}
\end{figure}
%--------------------------------------------------------------

%--------------------------------------------------------------
\subsection{Projecting to the spin-only model}
%--------------------------------------------------------------

After establishing the validity of the  $t$-$\varepsilon$-$U$-like level structure of Fig.~\ref{fig:energy_level} for 
the realistic parameters of the three-band Hubbard model with Zn impurity, 
we now turn to the low-energy properties of  the model (\ref{H:diagonalized_Cu}) and (\ref{H:diagonalized_Zn}).
To derive the low-energy model one needs to consider virtual transitions between different states.
As was argued above, since the system is at half-filling, the relevant transitions 
are between the lowest states from one- and two-hole sector of both Cu- and Zn-cluster states. 
Altogether, there are five different hopping integrals, as is shown in Fig.~\ref{fig:effHop}. 
For brevity, we switch to $t$'s for the hopping integrals from $F_{ii'}^{jj'}$ and $\tilde{F}^{jj'}_{ii'}$ in 
(\ref{H:diagonalized_Cu}) and (\ref{H:diagonalized_Zn}). For instance, $t_1=F_{10}^{01}$ is the hopping
of the spin to the neighboring empty site and $t_2=F_{11}^{20}$ is creating the Zhang-Rice singlet 
from the two one-hole sites leaving the other site empty.
For the transitions involving Zn-states, $t_{11}$ is between one-hole states on Cu and  Zn, while 
$t_{12}$ and $t_{21}$ are between the one-hole and two-hole states on Cu and Zn. 
The higher-energy transitions will be neglected as their contributions are small. 
The explicit expressions for hopping integrals can be obtained by evaluating matrix elements of the 
Hamiltonian in terms of the original Cu, Zn, and O operators, 
(\ref{H:symmetric_oxygen_Cu}) and (\ref{H:symmetric_oxygen_Zn}),  between the initial and final states in the basis 
of the local eigenstates $|\psi_l^{i(j)}\rangle$  and $ |\tilde{\psi}_\ell^{i(j)}\rangle$, see Appendix~\ref{AppA} for details.

 Next, we apply a canonical transformation to this extended $t$-$\varepsilon$-$U$-like model assuming
that the hopping integrals are smaller than the effective $U^{\rm Cu}_{\rm eff}$ and $\varepsilon_{\rm eff}^{\rm Zn}$. 
At half-filling and  in the second order, the $t$-$U$ part yields the Heisenberg model 
with $J = 4|t_2|^2/U_{\rm eff}^{\rm Cu}$ and with the negligible higher-order terms. 
In the presence of impurity, the same transformation  leads to the dilution-only model, 
in which the four adjacent spin-spin links are cut by the spinless Zn-site.
For the sites in the vicinity of impurities, we extend the transformation to the fourth
order\cite{Takahashi_77,Yoshioka_88,Chernyshev_04}  in $t/\varepsilon (U)$ 
to include the -$\varepsilon$- part of the model
 for the purpose of 
taking into account the effects of the in-gap impurity state.

 %--------------------------------------------------------------
\begin{figure}[t]
\includegraphics[width=1\columnwidth]{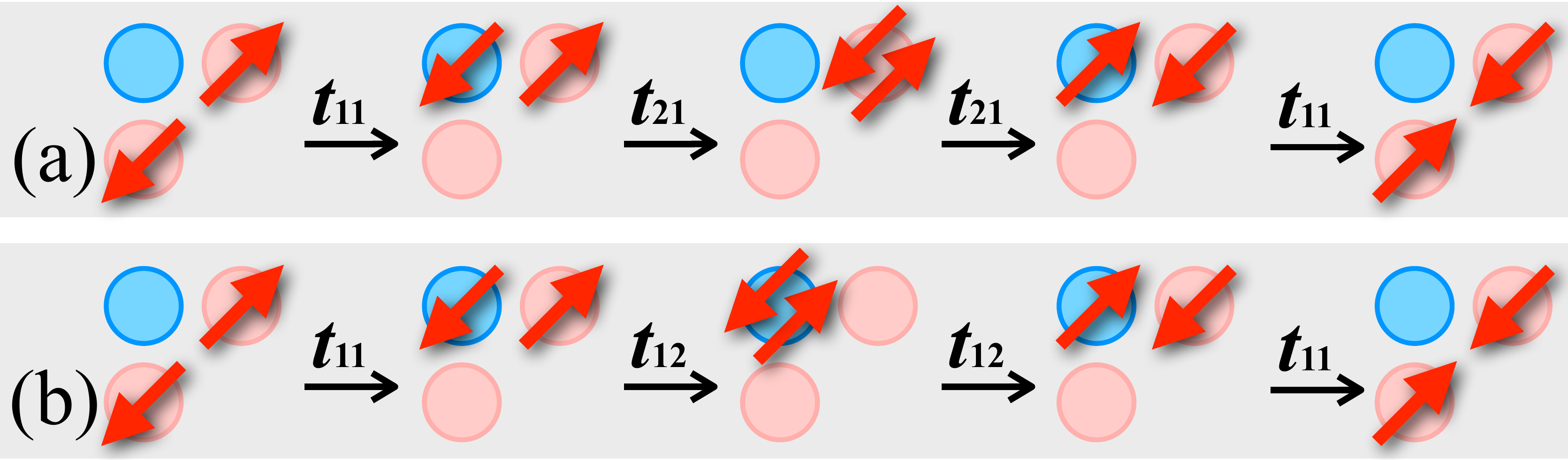}
\caption{(Color online) Superexchange processes for (a) $\delta J_1$, (b) $\delta J_2$. See text and Fig.~\ref{fig:effHop}
for notations.}
\label{fig:dJ12}
\end{figure}
%--------------------------------------------------------------
%--------------------------------------------------------------
\begin{figure}[t]
\includegraphics[width=1\columnwidth]{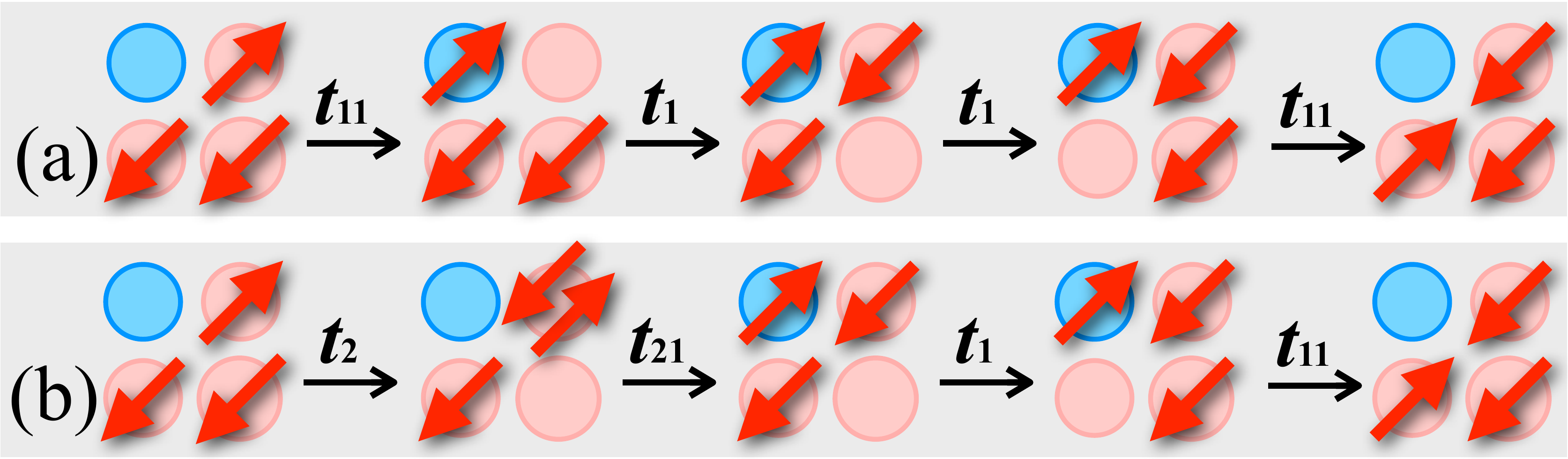}
\caption{(Color online) Ring-exchanges for  (a) $\delta J_3$, (b) $\delta J_4$.}
\label{fig:dJ34}
\end{figure}
%--------------------------------------------------------------
%--------------------------------------------------------------
\begin{figure}[b]
\includegraphics[width=1\columnwidth]{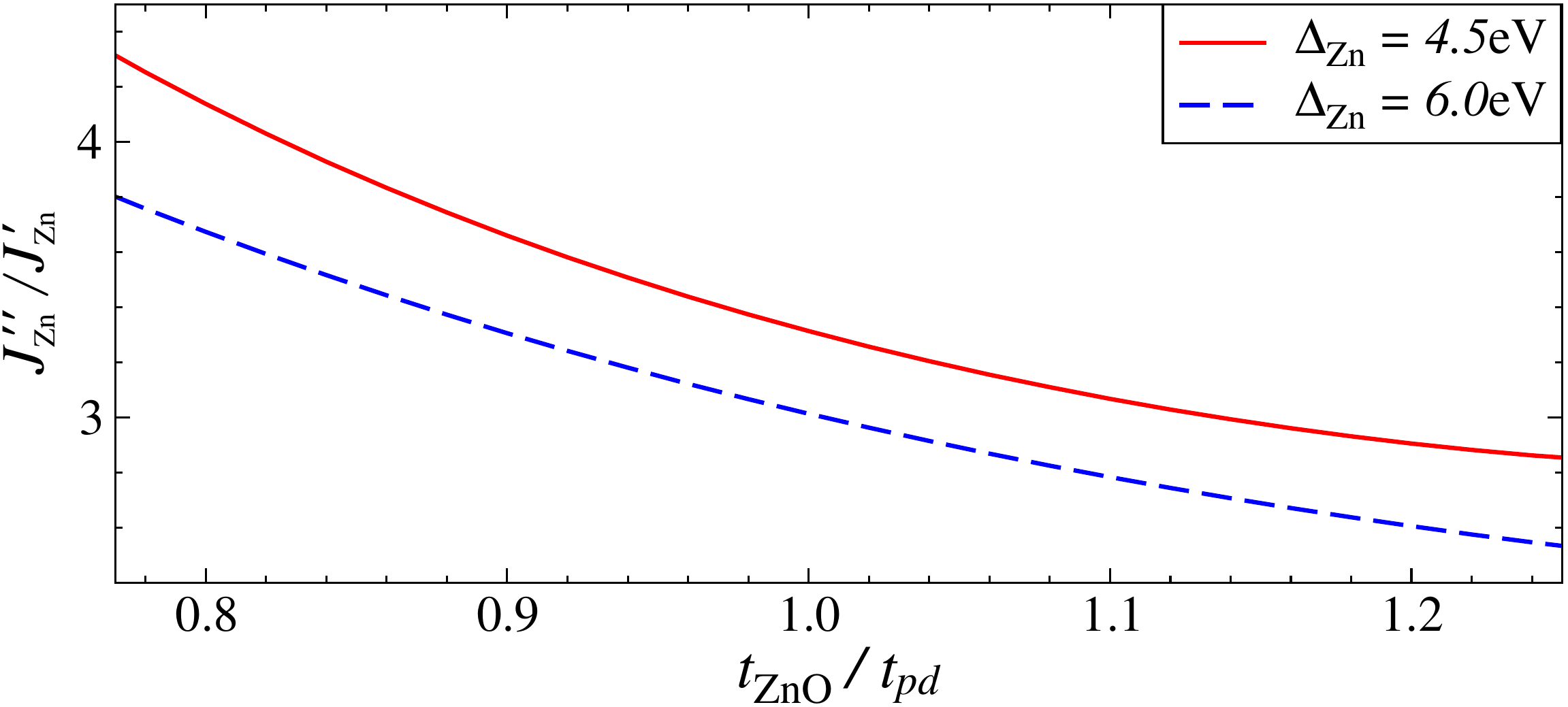}
\caption{(Color online) The ratio of $J_{\rm Zn}''/J_{\rm Zn}'$ vs $t_{\rm ZnO}/t_{pd}$ 
for two representative $\Delta_{\rm Zn}$. Other parameters are as in  Fig.~\ref{fig:eff_epsilon}.}
\label{fig:J3_J2}
\end{figure}
%--------------------------------------------------------------

In addition to deriving the low-energy, spin-only model with the parameters that can be traced to the three-band ones,
we are also able to analyze individual superexchange processes that contribute to the terms of that model.
We find that in the 4$^{\rm th}$ order there are two types of virtual transitions through the impurity site. 
First involves three sites, two Cu and one Zn, with the coppers either across the Zn-site or at the right angle,  as in 
Fig.~\ref{fig:dJ12}. Second type needs three Cu in addition to Zn impurity, arranged in a 4-site plaquette, see Fig.~\ref{fig:dJ34}.
Each of the transition types yields two superexchange channels shown in Figs.~\ref{fig:dJ12} and \ref{fig:dJ34}. First two, in 
Fig.~\ref{fig:dJ12}(a) and (b), are   the standard superexchanges, taking four virtual steps instead of the usual two.
We denote the corresponding couplings generated by such processes as $\delta J_1$ (doubly-occupied state on Cu)
and $\delta J_2$ (doubly-occupied state on Zn), respectively. The other two processes, in Fig.~\ref{fig:dJ34}(a) and (b),
are the ring exchanges, which we denote as $\delta J_3$ (no doubly-occupied state involved) and $\delta J_4$ 
(doubly-occupied state on Cu), respectively. These couplings are given by
\begin{eqnarray}
&\delta J_1 = \displaystyle\frac{4|t_{11}t_{21}|^2}{U_{\rm eff}^{\rm Cu}\big(\varepsilon_{\rm eff}^{\rm Zn}\big)^2},
\ \ &\delta J_2 = \frac{8|t_{11}t_{12}|^2}{\big(U_{\rm eff}^{\rm Zn}+2\varepsilon_{\rm eff}^{\rm Zn}\big)
\big(\varepsilon_{\rm eff}^{\rm Zn}\big)^2},
\label{Jis}\\
&\delta J_3 =- \displaystyle\frac{2|t_{11}t_{1}|^2}{\big(\varepsilon_{\rm eff}^{\rm Zn}\big)^3},
\ \ &\delta J_4 =- \frac{4|t_{11}t_{1}t_{2}t_{21}|}
{U_{\rm eff}^{\rm Cu}\big(\varepsilon_{\rm eff}^{\rm Zn}\big)^2},\nonumber
\end{eqnarray}
in terms of the notations of Fig.~\ref{fig:effHop}.
Assuming that  $\varepsilon^{\rm Zn}_{\rm eff}$ is smaller than $U_{\rm eff}$ according to the discussion of Sec.~\ref{Sec_II}.B, 
the magnitudes of $\delta J_i$'s should substantially exceed the conventional fourth-order terms of order 
$t_{\rm eff}^4/U_{\rm eff}^3$, which we neglect. Another important distinction of the couplings in (\ref{Jis}), is that they 
occur due to impurities and thus must have a direct relation to the doping-dependent effects.
Altogether, the considered processes combine into the next-nearest- and next-next-nearest-neighbor \emph{frustrating}
interactions around Zn-impurity, see Fig.~\ref{fig:JZnbonds}(c),
\begin{eqnarray}
&&J'_{\rm Zn} = \delta J_1 + \delta J_2 + \delta J_3 + \delta J_4,
\label{JZns}\\
&&J''_{\rm Zn} =  \delta J_1 + \delta J_2.\nonumber
\end{eqnarray}
It is clear that only superexchange-type processes of Fig.~\ref{fig:dJ12} can contribute to the next-next-nearest-neighbor
couplings across the impurity $J''_{\rm Zn}$, while the ring-exchanges of Fig.~\ref{fig:dJ34} are also at play for $J'_{\rm Zn}$.
Importantly, the latter terms have a ferromagnetic  sign, reducing the value of $J'_{\rm Zn}$ 
compared to $J''_{\rm Zn}$. Thus, counterintuitively,  the longer-ranged $J''_{\rm Zn}$ 
should be greater than $J'_{\rm Zn}$.
This feature is made explicit in our Fig.~\ref{fig:J3_J2}, which shows that the ratio of $J''_{\rm Zn}/J'_{\rm Zn}$ 
varies from about 4 to 2 for a representative range of the electronic parameters of Zn with the  other three-band 
model parameters as in Fig.~\ref{fig:eff_epsilon}.

\begin{figure}[t]
\includegraphics[width=1\columnwidth]{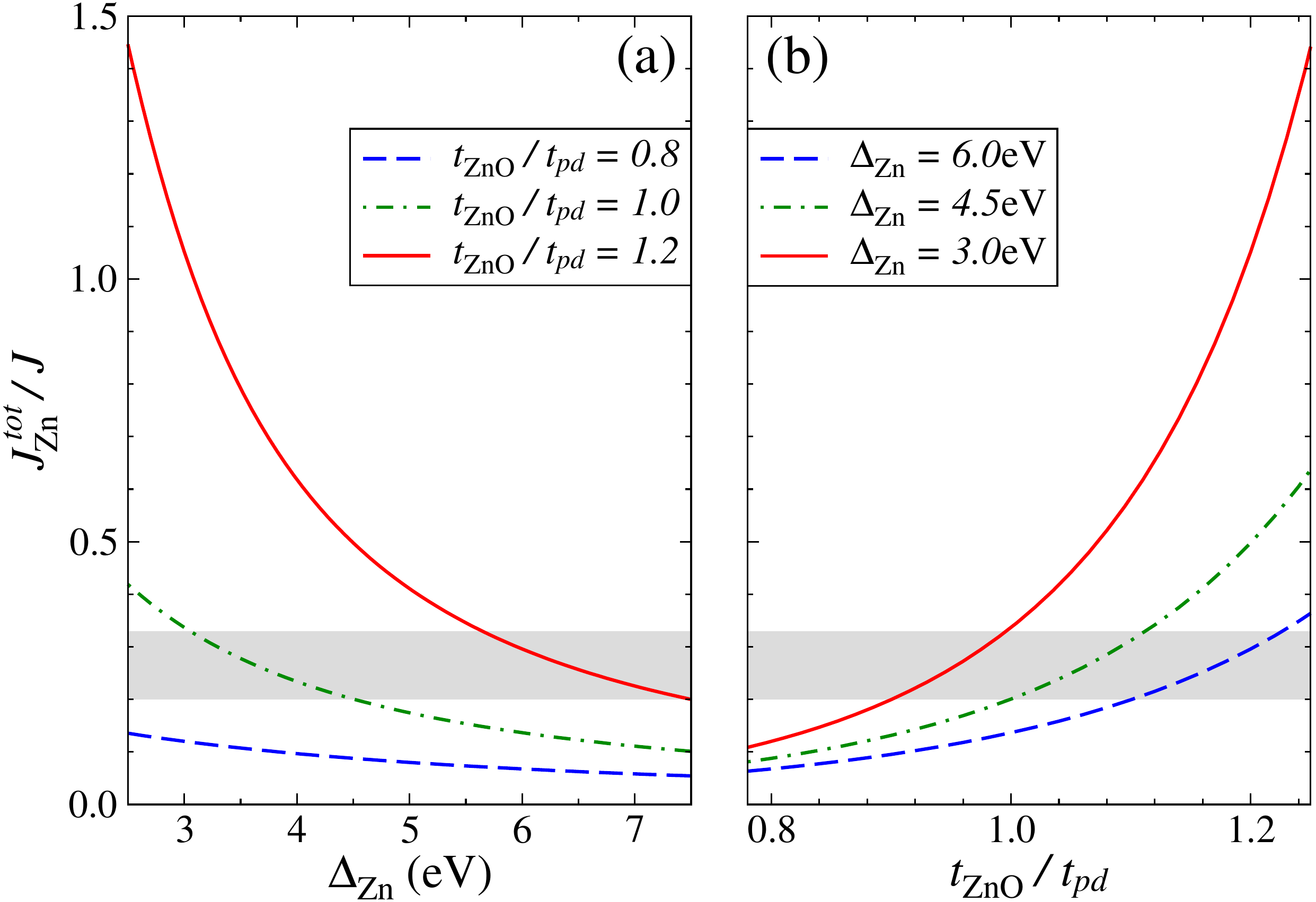}
\caption{(Color online) The total frustration effect per Zn impurity $J_{\rm Zn}^{tot} = 4J'_{\rm Zn}+2J''_{\rm Zn}$ 
(a) vs $\Delta_{\rm Zn}$  for several representative values of $t_{\rm ZnO}$ and (b) vs $t_{\rm ZnO}/t_{pd}$ for
several values of $\Delta_{\rm Zn}$, respectively. Three-band parameters are
$t_{pd}\!=\!1.5$eV, $\Delta\!=\!3$eV, $U_d\!\gg\!\Delta$ as before. 
The shaded area is the range of $J_{\rm Zn}^{tot}$ 
needed to explain experimental  reduction of the staggered 
magnetization $M(x)/M(0)$, see Sec.~\ref{Mx}. }
\label{fig:JznIratio}
\end{figure}

As is shown in Fig.~\ref{fig:JZnbonds}(c), each impurity generates four $J'_{\rm Zn}$ and two $J''_{\rm Zn}$. 
Thus, the total frustrating effect per Zn-impurity is $J_{\rm Zn}^{tot} = 4J'_{\rm Zn}+2J''_{\rm Zn}$. 
We show the dependence of $J_{\rm Zn}^{tot}$ on $\Delta_{\rm Zn}$  and  $t_{\rm ZnO}$ 
in Fig.~\ref{fig:JznIratio} with the same set of three-band parameters as 
in Sec.~\ref{Sec_II}.B and in Fig.~\ref{fig:eff_epsilon}. 
The shaded area in the graphs shows the range of $J_{\rm Zn}^{tot}$ 
needed to explain experimentally observed reduction of the staggered 
magnetization $M(x)/M(0)$, according to the $T$-matrix discussion of Sec.~\ref{Mx}. 
Clearly,  the uncertainty in the electronic parameters of  Zn-orbitals does not allow us to provide significant restrictions on the
value of $J_{\rm Zn}^{tot}$. However, the values needed for $M(x)/M(0)$ doping dependence outline the minimal requirements 
on such three-band model parameters. In particular, the coupling of Zn and O orbitals should not be much 
weaker than that of Cu and O, while the
position of the level on Zn is much less restricted. Overall, the realistic requirements on Zn electronic degrees of 
freedom leave a wide range of $J_{\rm Zn}^{tot}$ and they support the validity of our proposed impurity-induced
frustration mechanism.

%--------------------------------------------------------------
%--------------------------------------------------------------

Finally, the effective low-energy, spin-only model for La$_2$Cu$_{1-x}$Zn$_x$O$_4$ can be divided into two parts
\begin{equation}
{\cal H} =J \sum_{\langle ll'\rangle} \mathbf{S}_l \cdot \mathbf{S}_{l'}
+ J_{\rm Zn}'\sum_{\langle ll'\rangle'_\ell} \mathbf{S}_l \cdot \mathbf{S}_{l'}
 +J_{\rm Zn}''\sum_{\langle ll'\rangle''_\ell} \mathbf{S}_l \cdot \mathbf{S}_{l'} , \ \ \ 
\label{H:spin_only}
\end{equation}
the dilution-only model (first term), and the impurity-induced frustrating terms, 
with $\langle ll'\rangle'_\ell$ and $\langle ll'\rangle''_\ell$ denoting next- and next-next-nearest-neighbor bonds 
that are also nearest-neighbors of the impurity site $\ell$, see  Fig.~\ref{fig:JZnbonds}(c). The summations are 
carried over the Cu sites only.

We propose that the dilution-frustration model  (\ref{H:spin_only}) provides a proper description of 
Zn-doped CuO$_2$ planes and discuss its properties next.

%----------------------------------------------------------------------------------
\section{On-site magnetization in the low-energy model}
\label{Mx}
%----------------------------------------------------------------------------------

With the structure of the low-energy spin-only model for CuO$_2$ plane diluted by Zn impurities given
in (\ref{H:spin_only}), we investigate in detail the impact of impurities on
the order parameter, the average on-site magnetization $M$. 
We use the linear spin-wave approximation to the problem of single impurity in the 
model (\ref{H:spin_only}) and apply an analytical $T$-matrix approach to it, which  is based on summing up exactly 
infinite diagrammatic series for the impurity scattering amplitudes. Then we approximate the problem of finite
concentration of impurities as a linear superposition of scattering effects of individual random impurities, 
which amounts to an ``effective medium'' approach via disorder averaging. This step restores translational 
invariance for magnons but modifies their dispersion and provides them with damping through the impurity scattering 
self-energies. The averaging also takes into account the increase of the population of magnons around impurities, 
i.e., for the extra fluctuations of spins that reduce the value of the ordered moment. 
The latter effect is calculated from the renormalized 
magnon Green's function as a function of impurity concentration $x$.\cite{Chernyshev_01,Chernyshev_02}

\begin{figure}[t]
\includegraphics[width=1\columnwidth]{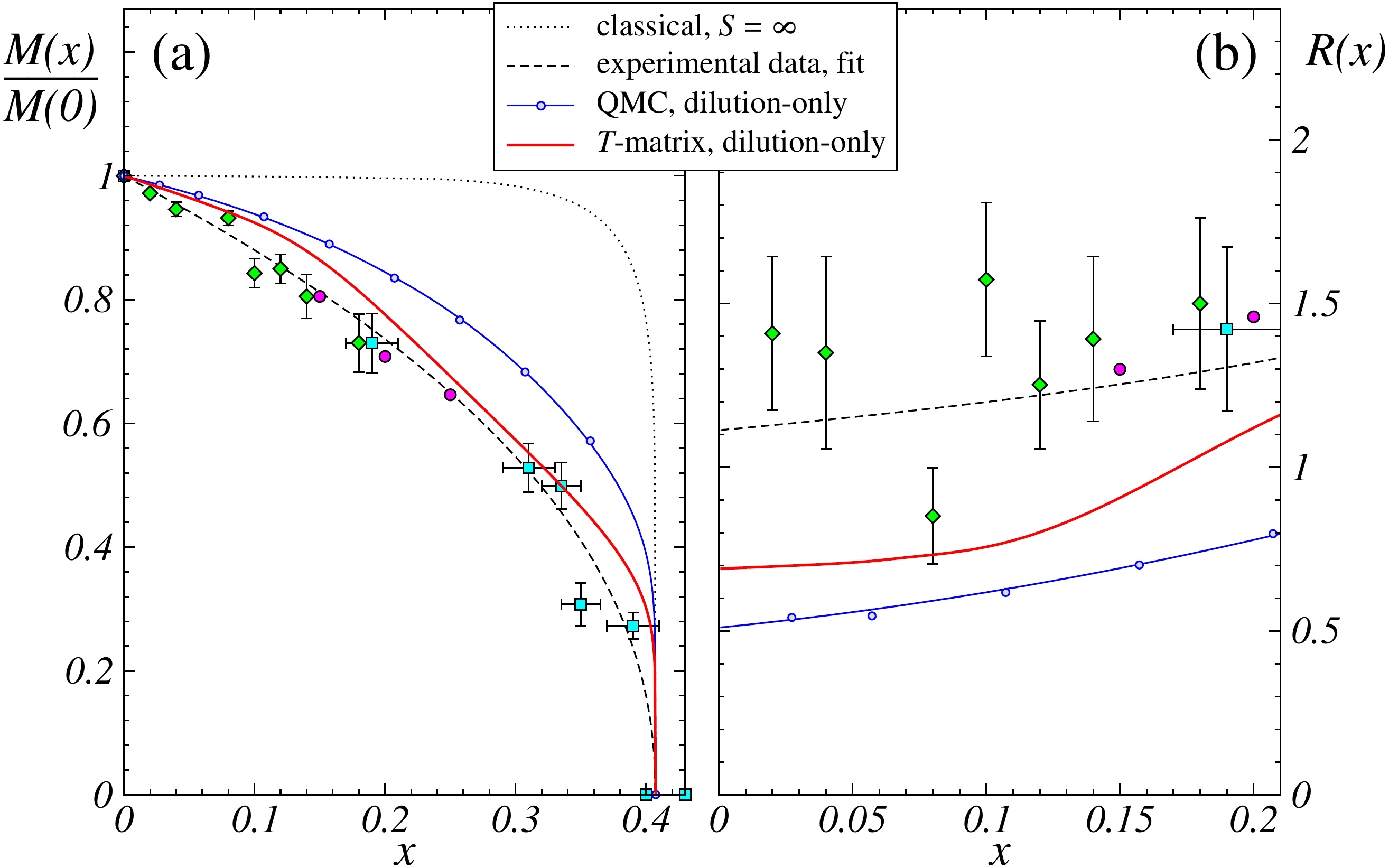}
\caption{(Color online) (a) 
The on-site magnetic moment $M(x)$ per magnetic site (Cu$^{2+}$), normalized to the undoped value $M(0)$
vs Zn doping $x$. Green diamonds and magenta circles are the NQR data,\cite{Corti_95,Carretta_97} 
cyan squares are the neutron scattering data,\cite{Greven_03}  and 
the dashed line is their best fit. The dotted line is the classical (Ising) result. Upper solid line (blue online) 
is the best fit of  the QMC results (blue circles)\cite {Sandvik02a} and the lower solid line (red online) is 
the $T$-matrix calculation results,\cite{Chernyshev_02} both for the dilution-only model. 
 (b) The slope function $R(x)$ of normalized staggered magnetization (see text) vs $x$. Symbols and lines 
 are the same as in (a).}
\label{fig:Mx_M0}
\end{figure}

We note that  the order parameter is normalized by the number of magnetic ions:
$M(x)=\sum_i|S_i^z|/N_m$, where $N_m=N-N_{imp}$ is 
the number of {\it magnetic} sites (Cu$^{2+}$) and $N$ is the total number of sites. 
The reason for that is twofold. First, the experimental data on $M(x)$ from such techniques 
as NMR[NQR] and neutron scattering are naturally normalized this way. Second, such a normalization also 
separates the classical effect of dilution from the purely quantum-mechanical suppression of the order parameter. 
For instance, $M(x)$ defined this way is  close to a constant for 
the classical (Ising) antiferromagnet on a square lattice,
as is shown in Fig.~\ref{fig:Mx_M0}(a), because this quantity is 
equivalent to a probability for a spin to belong to an infinite cluster,  
which is very close to 1 below the percolation threshold,
$x_p\approx 41\%$.  

In this Section, we first demonstrate the discrepancy of the previous theoretical works based on the
dilution-only model with experimental data and then proceed with the calculations within the dilution-frustration
model of Eq.~(\ref{H:spin_only}). Comparison with  experimental data provides a confirmation of the selfconsistency
of our consideration in two respects. First, it allows us to verify whether the impurity-induced frustration is at all a reasonable
mechanism of producing extra suppression of the order parameter. Second, we are able to compare the 
range of parameters $J'_{\rm Zn}$ and $J''_{\rm Zn}$ that are necessary to explain the experimental data for $M(x)$ 
with the range permitted by the three-band Hubbard mapping of Sec.~\ref{Sec_II}.

%-------------------------------------------------------------------------
%-------------------------------------------------------------------------

%--------------------------------------------------------------------- 
\subsection{Discrepancy with experiments}
%--------------------------------------------------------------------- 
 
Our Fig.~\ref{fig:Mx_M0}(a) shows a comparison of the results for the averaged on-site magnetic moment 
(staggered magnetization) $M(x)$, normalized to its value in the undoped system $M(0)$. 
The experimental data include  the NQR\cite{Corti_95,Carretta_97} 
and the neutron scattering data,\cite{Greven_03} with the dashed line being their best fit. 
Both set of theoretical results, from the QMC\cite {Sandvik02a} and the 
$T$-matrix\cite{Chernyshev_02} calculations, are for $T=0$  dilution-only model, i.e., neglecting frustrating effects
of impurities proposed in this work. 
First observation is that the unbiased QMC data agree very closely with the $T$-matrix results 
up to $x\simeq 15\%$, thus supporting the validity of the latter in the
low-doping regime.\cite{Chernyshev_02,Sandvik02a} At higher doping, single-impurity $T$-matrix
tends to overestimate the effect of impurities on the order parameter. 
Note that the $T$-matrix results in Fig.~\ref{fig:Mx_M0}(a) are multiplied by the 
classical probability (dotted line). This 
does not affect the data for $x<35\%$ but makes a comparison 
consistent  near the percolation threshold $x_p$.

However, there are substantial discrepancies between theoretical and experimental
results. In Fig.~\ref{fig:Mx_M0}(a) the experimental data are always below the
theoretical  curves for the dilution-only model. 
The difference is much more apparent in the slope function, defined as  
\begin{eqnarray}
R(x) = \frac{1}{x}\bigg( 1 - \frac{M(x)}{M(0)}\bigg)\, ,
\label{eq:slope_general}
\end{eqnarray}
and displayed in Fig.~\ref{fig:Mx_M0}(b). The quantity $R(x)$ has a transparent physical meaning at $x\rightarrow 0$:
it shows a reduction of the order parameter $M$ per impurity due to enhanced quantum fluctuations in the impurity's
neighborhood, the quantity that should be captured properly by the $T$-matrix approach.  Fig.~\ref{fig:Mx_M0}(b)
shows a large discrepancy of the experimental slope, $R^{\rm exp}(0)\approx 1.1$, with both the QMC, 
$R^{\rm QMC}(0)\approx 0.5$, a factor of approximately 2, and the $T$-matrix, $R^{T}(0)\approx 0.7$, 
a factor of about 1.6.  This gives a clear indication that the dilution-only theory significantly
underestimates the impact of the \emph{individual} impurities on the quantum spin background. 
Thus, the dilution-only model is not enough to describe La$_2$Cu$_{1-x}$Zn$_x$O$_4$.

%--------------------------------------------------------------------- 
\subsection{Qualitative $J_1^{\rm eff}$-$J_2^{\rm eff}$-$J_3^{\rm eff}$ mean-field consideration of 
the dilution-frustration model}
%--------------------------------------------------------------------- 

We would like to argue that the longer-ranged spin interactions in the \emph{undoped} system are 
unlikely to provide a resolution to the observed discrepancy in $M(x)$. On the other hand, the 
dilution-frustration model of Eq.~(\ref{H:spin_only}) can be approximated on a mean-field level as 
an effective model with further exchanges that are proportional to doping. 
Such a mean-field consideration offers a simple way of checking
the viability of this model in explaining the discrepancy.

One of the natural ideas to explain the disagreement is by including the longer-range  interactions, such as $J_2$, $J_3$, etc., 
and ring-exchange, in the low-energy spin-only model for the undoped CuO$_2$ plane. 
Such terms are generally present in any realistic low-energy spin models of Mott insulators and formally
come as a result of the higher-order expansion of the Hubbard-like models.\cite{Chernyshev_04}
Although for the cuprates\cite{Delannoy_09} such terms are of the order of 10\% of the nearest-neighbor $J$, 
they do lead to a reduction\cite{Tremblay_09} of $M$. 
Qualitatively, assuming the presence of $J_2$ term in the spin-model and using an expansion of $M$ in the 
dilution fraction $x$ and in $J_2$,  one can obtain on the mean-field level at small $x$ and $J_2$:
\begin{equation}
R(J_2)\approx
R(0)\left(1+A\frac{J_2}{J}\right),
\end{equation}
where the coefficient $A\alt 1$ follows from the consideration of the $J_1$-$J_2$ model\cite{Schulz_96} 
and $R(0)$ is the theoretical slope of the dilution-only 
model in Fig.~\ref{fig:frust_Mx_M0}(b). Thus, unless $J_2$ is of the same order as $J$, 
one can not expect a  large correction to the slope   of $M(x)/M(0)$ from the presence of further-neighbor
terms in the low-energy model.

In addition, since these terms suppress the order already in the undoped system, 
this mechanism is unlikely to enhance fluctuations specifically due to dilution. 
Put another way, dilution of the models with extended interactions breaks frustrated bonds
alongside the nearest-neighbor ones and may even induce more robust order around impurities. 
A recent detailed study\cite{Delannoy_09} has shown that while extended 
interactions such as $J_2$ and cyclic terms do lead to an overall lower {\it absolute} 
values of the on-site magnetization in the impurity-doped Hubbard model of La$_2$Cu$_{1-x}$Zn$_x$O$_4$, 
they do not significantly change the theoretical $M(x)/M(0)$ dependence and are not able to explain the 
large experimental slope $R(x)$ in Fig.~\ref{fig:frust_Mx_M0}(b).

On the other hand, the dilution-frustration model of Eq.~(\ref{H:spin_only}) offers an alternative:
while the undoped system can be considered as having no further-neighbor frustrating  terms in its low-energy model,
such terms are introduced by the  dopants. We thus propose an ``effective medium'', mean-field 
$J_1^{\rm eff}$-$J_2^{\rm eff}$-$J_3^{\rm eff}$ model, in which effective interactions depend on the doping concentration $x$,
as a simplified version of the dilution-frustration model (\ref{H:spin_only}) that allows for a straightforward calculation of
$M(x)$ without the complications of the $T$-matrix or QMC numerical approaches. While approximate, this 
method gives an intuitive picture of  the proposed mechanism and allows us to estimate whether it can be a viable source 
of enhancing  quantum fluctuations due to doping. 

In a sense, the mean-field approximation ``spreads''  the frustration provided by impurities evenly over the whole 
system. The strength of the effective couplings is related to the exchanges in  (\ref{H:spin_only}) by counting bonds:
nearest-neighbor interaction $J_1^{\rm eff} = J(1-2x)$, next-nearest-neighbor 
interaction $J_2^{\rm eff} = 2xJ_{\rm Zn}'$, and next-next-nearest-neighbor interaction 
$J_3^{\rm eff} = xJ_{\rm Zn}''$. Since we are interested in $x\ll 1$ limit, the derivation of  $M$ amounts
to on expansion of the spin-wave correction to the order parameter within the 
$J_1^{\rm eff}$-$J_2^{\rm eff}$-$J_3^{\rm eff}$ model to the first power in
$J_2^{\rm eff}/J$ and $J_3^{\rm eff}/J$. Using the approach of Ref.~\onlinecite{Schulz_96} we obtain
the moment reduction as
\begin{eqnarray}
\frac{M(x)}{M(0)} \simeq 1 - \frac{C_2}{M(0)}\left(\frac{J_2^{\rm eff}}{J}\right)
-\frac{C_3}{M(0)} \left(\frac{J_3^{\rm eff}}{J}\right)\, ,
\label{J2J3}
\end{eqnarray}
where $C_{2(3)}=\sum_{\bf k}\eta_{2(3),{\bf k}}\gamma_{\bf k}^2/2\omega^3_{\bf k}$, with 
$\omega_{\bf k}=\sqrt{1-\gamma_{\bf k}^2}$, $\gamma_{\bf k}=(\cos k_x +\cos k_y)/2$, 
$\eta_{2,{\bf k}}=(1-\cos k_x \cos k_y)$,  $\eta_{3,{\bf k}}=(2-\cos^2 k_x -\cos^2 k_y)$, and $M(0)=0.3034$ 
is the ordered moment of the $S=1/2$, square-lattice nearest-neighbor Heisenberg 
antiferromagnet in the spin-wave approximation. Interestingly, 
since the $C_{2}=0.2909$ and $C_{3}=0.5205$, the suppression of the order by $J_3^{\rm eff}$ is almost 
two times as effective as by $J_2^{\rm eff}$ of the same value.
Setting  the  change of the slope, which is needed for the dilution-only result
$R^{T}(0)\approx 0.7$ to achieve the agreement 
with experimental $R^{\rm exp}(0)\approx 1.1$, and using the relation of $J_{2}^{\rm eff}$ and $J_{3}^{\rm eff}$ to 
$J_{\rm Zn}'$ and $J_{\rm Zn}''$ we can determine the range of $J_{\rm Zn}^{tot} = 2 J_{\rm Zn}'' + 4 J_{\rm Zn}'$
necessary for that. Assuming   $J_2^{\rm eff} = J_3^{\rm eff}$ corresponds to the choice $J_{\rm Zn}''=2 J_{\rm Zn}'$, 
which is a reasonable ratio according to the  three-band model consideration, see Fig.~\ref{fig:J3_J2}. 
Using (\ref{J2J3}), the required total frustrating effect due to each impurity is  
$J_{\rm Zn}^{tot} \simeq 0.6 J$, which is well within the range of the estimates from the  three-band model mapping, 
see Fig.~\ref{fig:JznIratio}.

Altogether, this consideration shows that the dilution-frustration model is indeed a viable candidate for 
explaining the discrepancy with the experimental data.

%--------------------------------------------------------------------- 
\subsection{$T$-matrix for the dilution-frustration model}
%--------------------------------------------------------------------- 

A more rigorous, albeit more technically involved method of dealing with the dilution-frustration
model (\ref{H:spin_only}) is the $T$-matrix approach. Its advantage is in taking into account 
all multiple-scattering  magnon processes, by which it solves the single-impurity problem  exactly
within the $1/S$ spin-wave approximation. Since we are interested in identifying a mechanism of the 
order parameter suppression in the low-doping regime, such 
an approach should be able to provide if not the ultimate, but at least a quantitatively correct 
result.

The   $T$-matrix method has been used in the study of the dilution-only model
in the 2D square-lattice 
antiferromagnet.\cite{Bulut,kampf,Mahan,Wan,NN,Sandvik,KK_imp,Yasuda,Chernyshev_01,Chernyshev_02}
 Here we outline some of the basic steps needed 
to calculate staggered magnetization in the dilution-frustration model (\ref{H:spin_only}) 
and provide  necessary details in Appendix~\ref{AppB}.

Staggered magnetization at $T=0$ in the presence of impurities can be written as\cite{Chernyshev_02}
\begin{eqnarray}
M(x)= M(0) -\sum_{\bf k}\frac{1}{\omega_{\bf k}}
\left(\langle\alpha^\dag_{\bf k}\alpha^{\phantom{\dag}}_{\bf k}\rangle-
\gamma_{\bf k}\langle\alpha^\dag_{\bf k}\alpha^\dag_{\bf -k}\rangle\right)\, ,\ \ \
\label{eq:Mx}
\end{eqnarray}  
where $M(0)=S-\Delta \simeq 0.3034$ is the staggered magnetization of the $S=1/2$,  
square-lattice, nearest-neighbor Heisenberg 
antiferromagnet at zero-doping, which is already reduced by quantum 
fluctuations ($\Delta$) of the antiferromagnetic ground state, 
and $\omega_{\bf k}=\sqrt{1-\gamma_{\bf k}^2}$ as before. 
Quantum fluctuations due to impurities further reduce $M(x)$ by non-zero magnon expectation values 
$\langle\alpha^\dag_{\bf k}\alpha^{\phantom{\dag}}_{\bf k}\rangle$ and
$\langle\alpha^\dag_{\bf k}\alpha^\dag_{\bf -k}\rangle$.
These can be expressed through the  magnon Green's function,\cite{Chernyshev_02} leading to
\begin{eqnarray}
\frac{M(x)}{M(0)} &=& 1 -\frac{1}{M(0)}\sum_{\bf k}\int_{-\infty}^{0} \frac{d\omega}{\pi\omega_{\bf k}}
\Big({\rm Im}G_{\bf k}^{11}(\omega) 
\nonumber \\
&&\phantom{-\frac{1}{M(0)}\sum_{\bf k}\int_{-\infty}^{0}\frac{d\omega}{\pi\omega_{\bf k}}}
-\gamma_{\bf k}{\rm Im}G_{\bf k}^{12}(\omega)\Big)\, , \ \ 
\label{eq:Mx_M0}
\end{eqnarray}  
where $G_{\bf k}^{11}$ and $G_{\bf k}^{12}$ are the diagonal and the off-diagonal component of the $2\times 2$ 
matrix Green's function, respectively, see Eq.~(\ref{eq:Greens_function}) for the corresponding Dyson-Belyaev expression of
them via the magnon self-energies. We note that ${\rm Im}G_{\bf k}^{11}(\omega)>0$ for $\omega<0$.
Magnon self-energy  in this approach comes as a result of averaging over  random impurity distribution, 
which relates it to the forward-scattering (${\bf k}'={\bf k}$) elements of the $T$-matrix
\begin{eqnarray}
\mathbf{\hat{\Sigma}}_{\bf k}(\omega) = 
x\sum_{\mu , I}\delta_{{\bf k},{\bf k}'} \mathbf{\hat{T}}^{I\mu}_{{\bf k},{\bf k}'}(\omega)\, .
\label{eq:self_energy}
\end{eqnarray}
Here the sum includes the contributions from impurities in both sublattices  ($I = A$, $B$) 
and from each of the non-zero $\mu = s$-, $p$-, and $d$-wave components of the $T$-matrix for the vacancy in the 
nearest-neighbor square lattice.
Clearly, the self-energies are also proportional to the impurity concentration $x$.

The individual components of the $T$-matrix (\ref{eq:Tmatrices}) obey linear integral equations 
for the multiple-scattering in each partial wave. 
Such equations contain separable scattering potentials, which reduce integral equations to simple algebraic ones and
allow for solutions in a separable form. That is, each partial wave $T$-matrix  
$\mathbf{\hat{T}}^{I\mu}_{{\bf k},{\bf k}'}(\omega)$ is given by the product of the ${\bf k}$- and ${\bf k}'$-dependent 
scattering potential  components with the corresponding
$\omega$-dependent resolvents,\cite{Chernyshev_02}
 see Eqs.~(\ref{eq:Tmatrices}), (\ref{eq:Vs}), (\ref{eq:wave_vectors_A}), 
and  (\ref{eq:gamma_functions}) in Appendix~\ref{AppB}.
 
The dilution-frustration model (\ref{H:spin_only}) contains extra frustrating antiferromagnetic 
terms compared to the dilution-only model, $J_{\rm Zn}'$ and $J_{\rm Zn}''$, 
that couple spins around impurities.
We find that because these extra terms connect spins that belong to the same antifferomagnetic sublattice of the host, 
additional scattering provided by them is orthogonal to the $s$-wave due to 
a cancellation between $S^zS^z$ and $S^+S^-$ contributions.
For the same reason,  the 
next-next-nearest-neighbor  interaction $J_{\rm Zn}''$ also does not modify
the $d$-wave component of the scattering potential and contributes only to the $p$-wave, while the 
next-nearest-neighbor $J_{\rm Zn}'$ interaction 
affects  both the $d$-wave and the $p$-wave scattering, see (\ref{eq:Vs}) and (\ref{eq:Cs}). 

While the $x$-dependence of Eq.~(\ref{eq:Mx_M0}) obviously extends beyond linear, it is the linear term which
is of primary importance as it is defining the initial slope of the normalized magnetization, $R(0)$ in (\ref{eq:slope_general}).  
It is also the term that is expected to be treated properly by the $T$-matrix approach. 
Thus, one may want to simplify the general expression (\ref{eq:Mx_M0}) and obtain  an expression which is
explicitly linear in $x$.
Having in mind that all magnon self-energies in (\ref{eq:self_energy}) are  $\propto x$,
we use the Green's function expansion in $\hat{\Sigma}$, see (\ref{eq:Greens_function_elements}), to obtain
\begin{eqnarray}
&&\frac{M(x)}{M(0)}\simeq 1 - xR(0)
= 1 - \frac{1}{M(0)}\sum_{\bf k}\Bigg\{ 
\frac{-\gamma_{\bf k}{\rm Re}\Sigma_{\bf k}^{12}(\omega_{\bf k})}{2\omega_{\bf k}^2}
\nonumber\\
&&\phantom{\frac{M(x)}{M(0)}} +\int_{-\infty}^0\frac{d\omega}{\pi\omega_{\bf k}}\left[ 
\frac{{\rm Im}\Sigma_{\bf k}^{11}(\omega)}{\big(\omega - \omega_{\bf k}\big)^2}
  +\frac{\gamma_{\bf k}{\rm Im}\Sigma_{\bf k}^{12}(\omega)}
{\big(\omega^2 - \omega_{\bf k}^2\big)} \right]
 \Bigg\},
 \label{eq:magnetization_expand}
\end{eqnarray}
where we note again that ${\rm Im}\Sigma_{\bf k}^{11}(\omega)>0$ for $\omega<0$.
With this, we can study the effect of impurity-induced frustration on the order parameter by 
calculating integrals in (\ref{eq:magnetization_expand}) and (\ref{eq:Mx_M0}) for different values of
$J_{\rm Zn}'$ and $J_{\rm Zn}''$.

First, in agreement with the qualitative consideration provided by  
the mean-field $J_1^{\rm eff}$-$J_2^{\rm eff}$-$J_3^{\rm eff}$ model in 
Sec.~\ref{Mx}.B, we find that the next-next-nearest neighbor $J^{\prime\prime}_{\rm Zn}$   bond 
suppresses the order about as effectively as two next-nearest $J^{\prime}_{\rm Zn}$ bonds 
of the same strength. As was discussed in our previous work,\cite{Liu_09} 
the same result has also been obtained by the QMC calculations, confirming again a good qualitative and quantitative 
accord between the analytical $T$-matrix approach and an unbiased numerical method.
Since the three-band model calculations of Sec.~\ref{Sec_II} also suggest that 
the $J^{\prime\prime}_{\rm Zn}$-term is systematically larger than the 
$J^{\prime}_{\rm Zn}$-term, these findings make the $J^{\prime\prime}_{\rm Zn}$  interaction
particularly important.  

%--------------------------------------------------------------------- 
\begin{figure}[t]
\includegraphics[width=1\columnwidth]{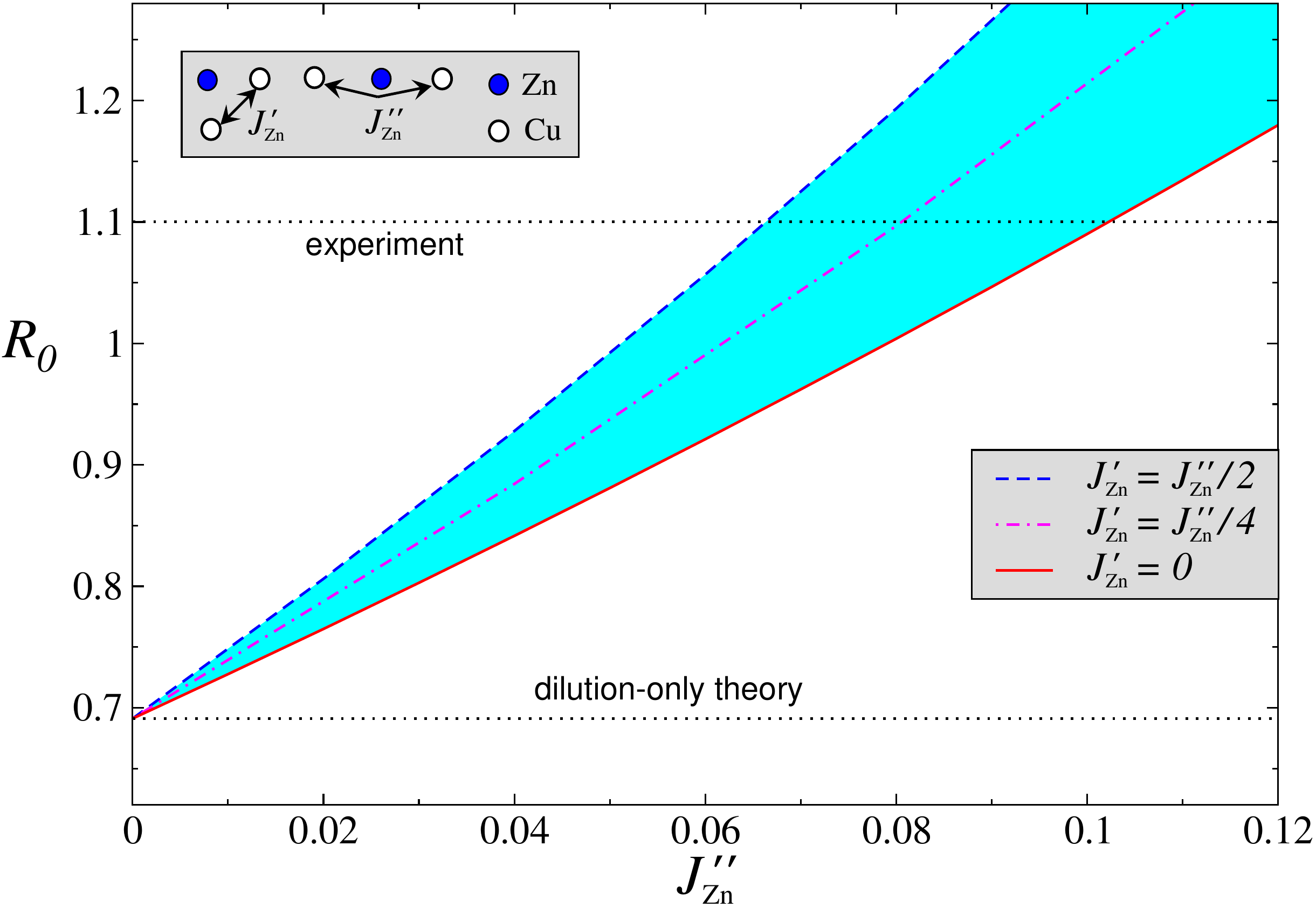}
\caption{(Color online) $R(0)$ from (\ref{eq:magnetization_expand}) 
vs $J_{\rm Zn}''$ (in  units of $J$). The red solid, magenta dotted-dashed, and blue dashed lines are 
for $J^{\prime}_{\rm Zn}/J^{\prime\prime}_{\rm Zn}=0$, $1/4$ and $1/2$, respectively. 
Colored area highlights the range of $J_{\rm Zn}''$.
The  experimental value $R^{exp}(0)$ and the dilution-only $T$-matrix result\cite{Chernyshev_02} 
of $R^T(0)$ are shown by the dotted horizontal lines.}
\label{fig:frust_R0}
\end{figure}
%--------------------------------------------------------------------- 

We study the  rate of suppression of the order parameter $R(0)$ in (\ref{eq:magnetization_expand}) 
for several representative ratios between $J^{\prime}_{\rm Zn}$ and $J^{\prime\prime}_{\rm Zn}$
as a function of one of them. Our Fig.~\ref{fig:frust_R0} shows $R(0)$ vs $J^{\prime\prime}_{\rm Zn}$
for three choices of $J^{\prime}_{\rm Zn}/J^{\prime\prime}_{\rm Zn}=0$, $1/4$ and $1/2$.
The dilution-only $T$-matrix result\cite{Chernyshev_02} of $R^T(0)\approx 0.69$ and the 
``target'' experimental value $R^{exp}(0)\approx 1.1$ are shown by the dotted horizontal lines.
We find the effect of frustrating interactions to be about two times stronger than in the mean-field
$J_1^{\rm eff}$-$J_2^{\rm eff}$-$J_3^{\rm eff}$ consideration of  Sec.~\ref{Mx}.B, implying that the averaged 
mean-field
treatment underestimates the effect of the order suppression due to frustration by local defects.
This result also means that the necessary  frustration is smaller than the one estimated from the
$J_1^{\rm eff}$-$J_2^{\rm eff}$-$J_3^{\rm eff}$ model.
 
One can see that the experimental value of $R(0)$
in Zn-doped La$_2$CuO$_4$ is met by the $T$-matrix results 
of the dilution-frustration model  at $J^{\prime\prime}_{\rm Zn}\approx 0.07J$ if the ratio of 
$J^{\prime}_{\rm Zn}/J^{\prime\prime}_{\rm Zn}$ is fixed to
$1/2$ and at $J^{\prime\prime}_{\rm Zn}\approx 0.08J$
for $J^{\prime}_{\rm Zn}/J^{\prime\prime}_{\rm Zn}=1/4$, the window of variation of 
$J^{\prime}_{\rm Zn}/J^{\prime\prime}_{\rm Zn}$ suggested by the three-band model consideration in Fig.~\ref{fig:JznIratio}
of Sec.~\ref{Sec_II}. These values translate into the total frustrating effect of impurity $J_{\rm Zn}^{tot}\approx 0.28J$
and $J_{\rm Zn}^{tot}\approx 0.24J$, respectively, which 
are well within the window suggested by the  three-band model calculations.  The somewhat wider 
range is highlighted by the gray shaded area in Fig.~\ref{fig:JznIratio}(a) and (b).

\begin{figure}[t]
\includegraphics[width=1\columnwidth]{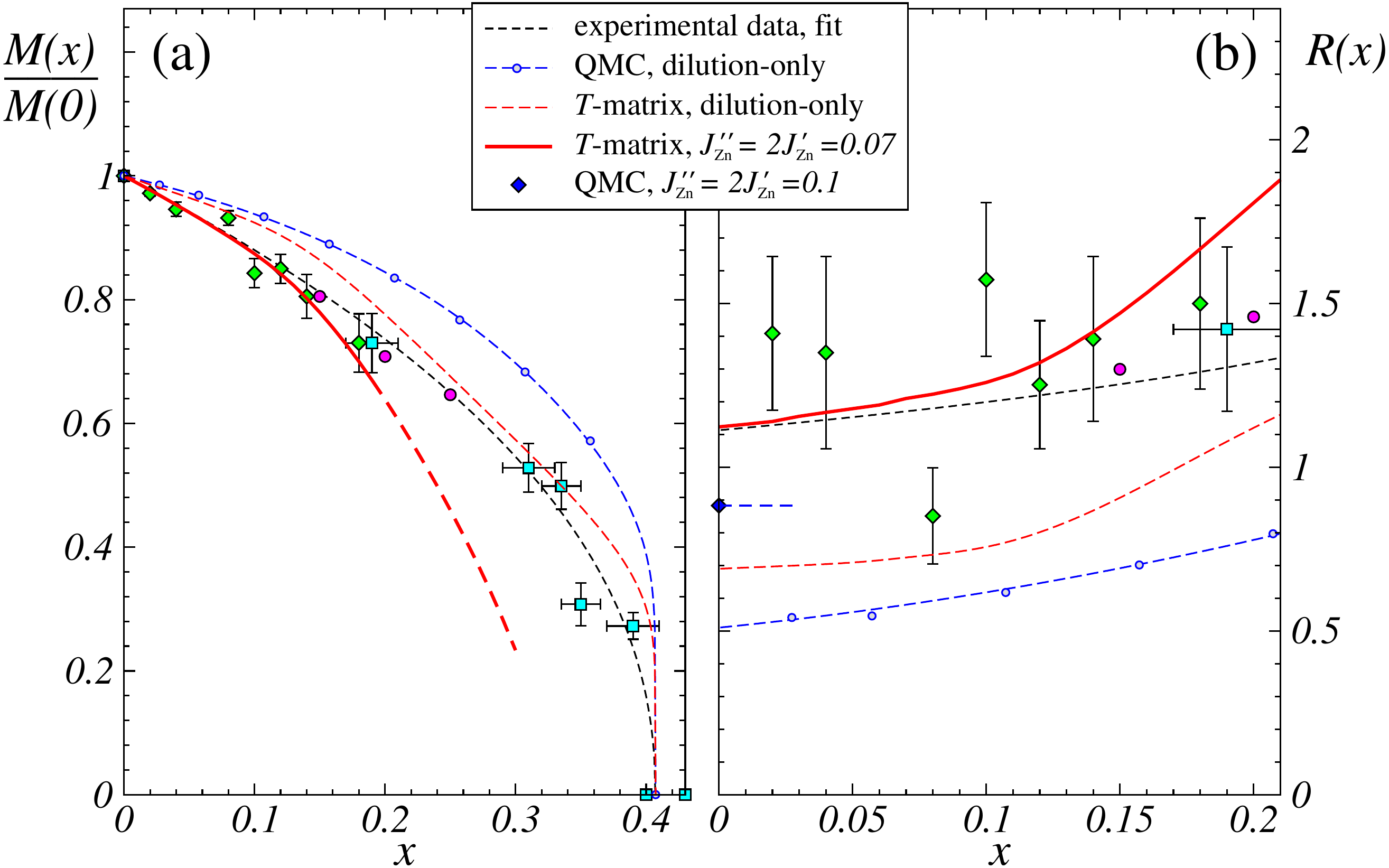}
\caption{(Color online) Same as Fig.~\ref{fig:Mx_M0} with 
the lines used for the fit of the QMC data and for  the $T$-matrix results of the dilution-only case  
switched to the dashed lines for clarity. 
The red solid line (and its long-dashed line tail) is the $T$-matrix results (\ref{eq:Mx_M0}) 
for the dilution-frustration model with $J_{\rm Zn}''=2 J_{\rm Zn}' = 0.07J$.
In (b) the blue diamond  with the dashed line is the QMC 
result for $J_{\rm Zn}''=2 J_{\rm Zn}' = 0.1J$ from Ref.~\onlinecite{Liu_09}.}
\label{fig:frust_Mx_M0}
\end{figure}

In our previous work,\cite{Liu_09}
QMC results seem to suggest a  higher value $J_{\rm Zn}^{tot}\!\agt\!0.4J$, 
still a modest amount of frustration, well within the range permitted by the three-band model. 
An example of the QMC results for $R^{QMC}(0)$ for the choice of parameters
$J^{\prime\prime}_{\rm Zn}\!=\!2J^{\prime}_{\rm Zn}\!=\!0.1J$ is shown in Fig.~\ref{fig:frust_Mx_M0}(b)  by the
blue diamond on the vertical axis. However, the QMC calculations may  be affected by the frustrating
nature of the impurity-induced interactions, which is associated with the infamous sign problem. This is also restricting
the use of the QMC  for the dilution-frustration model to the single-impurity problem.
In addition, the QMC results for $R^{QMC}(0)$ are obtained from the finite-size 
scaling that may overestimate $J_{\rm Zn}^{tot}$. Thus, using the data only from the largest clusters,\cite{Liu_09}
extrapolation for the 
$J^{\prime\prime}_{Zn}\!=\!2J^{\prime}_{Zn}\!=\!0.1J$ data set
gives $R^{QMC}(0)\!\approx\!1.0$, much closer to the experimental value. 
Thus,   the QMC result also demonstrates that the impurity-induced frustrations 
affect the staggered magnetization significantly and bring  the slope much closer to experimental data.

%--------------------------------------------------------------------- 
%--------------------------------------------------------------------- 

Using Eq.~(\ref{eq:Mx_M0})  we calculate  the normalized staggered magnetization and the slope 
(\ref{eq:slope_general}) for the dilution-frustration model with 
$J_{\rm Zn}'' = 2 J_{\rm Zn}' = 0.07J$ as a function of $x$, 
with the results shown in Fig.~\ref{fig:frust_Mx_M0}(a) and (b) by the solid lines. 
The comparison is provided with the experimental data and the theoretical $T$-matrix and QMC calculations 
for the dilution-only model  from Fig.~\ref{fig:Mx_M0}. 
With the frustrating parameters chosen to match the initial slope of the experimental best fit, 
obviously,  $M(x)/M(0)$ in the dilution-frustration model 
agrees much better with experiments. At higher values of $x\agt 15\%$, the $T$-matrix overestimates
the effect of impurities on $M(x)$ as it neglects the multiple-impurity scattering effects, similarly to the dilution-only case. 

Altogether, all of the evidences indicate that the dilution-frustration model describes La$_2$Cu$_{1-x}$Zn$_x$O$_4$ 
better than the dilution-only model and that the effective frustrating interactions via available electronic states of 
the impurity should be taken into account when studying doped Mott- and charge-transfer insulators.

%--------------------------------------------------------------------- 
\subsection{Further developments}
%--------------------------------------------------------------------- 

Based on the analysis of the N\'{e}el temperature doping dependence within the 
classical Heisenberg model, an alternative suggestion has been put forward for 
additional interactions introduced by dopants, such as Mg and Zn:
lattice distortions in the vicinity of the impurity site change the
strength of $J$ bonds.\cite{Oguchi}
The impact of this mechanism on $M(x)$ and its possible viability as an alternative to our proposal
 has been investigated using the QMC analysis in Ref.~\onlinecite{LiuKao10}. It was found 
 that changing the strength of the eight Cu-Cu bonds in the immediate vicinity of Zn site by as much as 
 15\% ($\equiv\delta \widetilde{J}_{tot}=1.2J$)  changes the slope of $M(x)/M(0)$ by at most 
a few percent. Such a weak effect rules out the lattice-distortion mechanism 
as a viable alternative to our theory.

One of the natural consequences of our idea is that different isovalent impurities, such as Zn$^{2+}$ and Mg$^{2+}$, 
should induce different amount of quantum fluctuations around them 
due to differences in their electronic levels available to generate frustrating interactions, 
and thus lead to a different rate of
suppression of the order parameter. Specifically,  
Mg$^{2+}$ should have no levels in any reasonable proximity
to the chemical potential, as opposed to Zn$^{2+}$, 
and thus should not lead to any substantial frustration, suggesting that the Mg-doped
 La$_2$CuO$_4$ must be described much better by a simple dilution-only model. 
This scenario has been investigated recently by the $\mu$SR  experiments in Zn- and  Mg-doped 
La$_2$CuO$_4$ at low doping.\cite{Carretta_11} 
The crucial finding of this work is that the spin stiffness, determined from the N\'{e}el temperature,
shows a stronger suppression rate in the case of Zn-doping, 
in agreement with the expectations from our theory. 
Although in additional to quantum contributions the slope  of $T_N(x)/T_N(0)\approx 1-\alpha x$ 
contains significant classical terms as well as logarithmic contributions from the 
three-dimensional interplane couplings,\cite{Chernyshev_02,Oguchi}
the difference in such slopes found in Ref.~\onlinecite{Carretta_11} is significant: $\alpha_{\rm Mg}\approx 2.7$ vs
$\alpha_{\rm Zn}\approx 3.5$. These values  are consistent with the difference expected 
between the dilution-only model\cite{Chernyshev_02}
and the dilution-frustration model with the parameters discussed in Sec.~\ref{Mx}.C.

%--------------------------------------------------------------------- 
\section{conclusions}
%--------------------------------------------------------------------- 

In this work, we have provided a detailed derivation of the low-energy, spin-only model of the
Zn-doped La$_2$CuO$_4$ starting from the realistic 
site-diluted three-band Hubbard model. We have followed with an equally detailed analysis of the order 
parameter in this low-energy model.
We have elaborated on the proposal of our previous work that the impurities in strongly
correlated systems may induce significant longer-range frustrating interactions among the spins, 
which are absent or negligible in the corresponding undoped system, 
due to hybridized electronic states of the impurity at the scale less than the Hubbard-$U$.
Such  impurity-induced frustrating interactions are shown   to significantly enhance local 
quantum fluctuations compared to the dilution-only model. In particular, the 
dilution-frustration model demonstrates stronger suppression of the order and, for a 
choice of parameters appropriate for Zn-doped La$_2$CuO$_4$, it resolves
discrepancies between   experiments and earlier theories.  
Recently, our theory has received further experimental 
confirmation from the $\mu$SR studies of   spin stiffness in Zn and Mg doped La$_2$CuO$_4$. 

Although our work considers a particular case of diluted La$_2$CuO$_4$, this study 
has far-reaching consequences for other diluted antiferromagnets and doped Mott and charge-transfer 
insulators.\cite{Zaliznyak}
One of the intriguing consequences of the proposed mechanism is the change of the character 
of the percolation transition due to frustrating interactions across the impurities. The impurity doping of 
one-dimensional spin systems such as  spin chains and ladders should introduce weaker links between their parts
instead of breaking them into independent pieces. 
Experiments in diluted frustrated spin systems is yet another area, such as recently studied diluted $J_1$-$J_2$ 
system,\cite{Carretta_05,Carretta_12} which should also be affected by the same mechanism. 
Another perspective is offered by an extension of the proposed mechanism of the impurity-induced frustrating 
interactions to the case of the doping away from half-filling where it may be 
responsible for the pair-breaking mechanism in the doped 
CuO$_2$ planes.

%----------------------------------------------------------------------------
\begin{acknowledgments}

We would like to thank Cheng-Wei Liu, Ying-Jer Kao, and Anders Sandvik for collaboration on the earlier 
work and for numerous conversations that has shaped our current understanding of the problem. 
We are thankful to Pietro Carretta for many useful communications and for sharing his results prior to publication.
We acknowledge useful discussions with
A.~Balatsky,
M.~Gingras,
M.~Greven,
P.~Hirschfield,
P.~Holdsworth,
I.~Martin,
N.~M.~Plakida, 
A.-M.~Tremblay,
and I.~Zaliznyak.

 This work was supported by DOE under grant
DE-FG02-04ER46174. 
Part of this work has been done at the Kavli Institute of Theoretical Physics. 
The research at KITP was supported by the NSF under Grant No.~PHY11-25915.

\end{acknowledgments}
%----------------------------------------------------------------------------

%----------------------------------------------------------------------------
\appendix
%----------------------------------------------------------------------------

%------------------------------------------------------------------
\section{Details of the three-band model consideration}
\label{AppA}
%------------------------------------------------------------------
%------------------------------------------------------------------
\subsection{Wannier orthogonalization of oxygen orbitals}
%------------------------------------------------------------------

The hopping terms in the three-band model (\ref{H:three-band-Cu}) and (\ref{H:three-band-Zn}) couple the Cu and Zn
states with the symmetric combination of O-states, i.e.,
\begin{eqnarray} 
{\cal H}_{pd} =  -t_{pd}\sum_{l\alpha}\left(d^\dagger_{l\alpha}P^{\phantom\dagger}_{l\alpha}+{\rm H.c.}\right)\, ,
\label{H:tpd}
\end{eqnarray}
with 
\begin{eqnarray} 
P^{\phantom\dagger}_{l\alpha}=\left(p^x_{l+\frac{{\bf x}}{2}}
+p^x_{l-\frac{{\bf x}}{2}}+p^y_{l+\frac{{\bf y}}{2}}+p^y_{l-\frac{{\bf y}}{2}}\right)_\alpha\, ,
\label{p_sym}
\end{eqnarray}
where  ${\bf x}$ and ${\bf y}$ are the vectors in the 
$x$ and $y$ direction of length $a_0=1$, the CuO$_2$ lattice spacing.
The problem is that the operators $P_{l\alpha}$ do not obey proper anticommutation relations with the ones on the 
nearest-neighbor CuO$_4$(ZnO$_4$) clusters, because one of the four $p$-operators in (\ref{p_sym}) belongs to both clusters. 
The Wannier-orthogonalization 
procedure is performed in the ${\bf k}$-space.\cite{ZR,Chernyshev_94_01} 
Since the O-lattice is fully periodic even for the CuO$_2$ plane doped with Zn, the 
Fourier transform of the oxygen-hole operators $p_m^{x(y)}$ can be written as
\begin{eqnarray}
&&p^{x(y)}_{m\alpha} =\frac{1}{\sqrt{N}}\sum_{\bf k} e^{-i{\bf k}\cdot\left({\bf r}_l\pm
\frac{{\bf x}}{2}\left(\frac{{\bf y}}{2}\right)\right)} \, p^{x(y)}_{{\bf k},\alpha},
\label{eq:pk}
\end{eqnarray}
where the sum is over the Brillouin zone and  $N$ is the number of unit cells. 

Motivated by (\ref{p_sym}) and using (\ref{eq:pk}), one can introduce the orthogonalized symmetric 
and antisymmetric oxygen operators  in momentum space as
\begin{eqnarray}
&&q_{{\bf k},\alpha}=\lambda_{\bf k}^{-1}\bigg[\cos\left(\frac{k_y}{2}\right)p^y_{{\bf k},\alpha}+
\cos\left(\frac{k_x}{2}\right)p^x_{{\bf k},\alpha}\bigg],
\label{eq:qs}\\
&&\tilde{q}_{{\bf k},\alpha}=\lambda_{\bf k}^{-1}\bigg[\cos\left(\frac{k_x}{2}\right)p^y_{{\bf k},\alpha}-
\cos\left(\frac{k_y}{2}\right)p^x_{{\bf k},\alpha}\bigg],\nonumber
\end{eqnarray}
with the normalization factor 
\begin{eqnarray}
\lambda_{\bf k}=\sqrt{1+\gamma_{\bf k}},
\end{eqnarray}
and $\gamma_{\bf k}=\frac{1}{2}\left(\cos{k_x}+\cos{k_y}\right)$. These operators are now properly normalized 
and obey regular fermionic anticommutation relations.

Then, using 
\begin{eqnarray} 
P_{l\alpha}=\frac{1}{\sqrt{N}}\sum_{\bf k} e^{-i{\bf k}\cdot{\bf r}_l} \lambda_{\bf k} \, q_{{\bf k},\alpha}\, ,
\label{p_sym_k}
\end{eqnarray}
and applying the inverse Fourier transform yields the hopping Hamiltonians in (\ref{H:symmetric_oxygen_Cu}) and
(\ref{H:symmetric_oxygen_Zn}) with the real-space Wannier amplitudes given by
\begin{eqnarray}
&&\lambda_{ll'} = \sum_{{\bf k}}\lambda_{{\bf k}}e^{-i{\bf k}\cdot({\bf r}_l - {\bf r}_{l'})}\label{eq:lambda_ll'}\, ,
\end{eqnarray} 
which depend on the distance between the clusters. 
As a result, the original three-band Hamiltonian (\ref{H:three-band-Cu}) with Zn-impurity (\ref{H:three-band-Zn})
is rewritten in a much more convenient, symmetric CuO$_{4}$(ZnO$_4$) cluster form, containing oxygen degrees of freedom 
that couple to copper and zinc orbitals most effectively.  On the other hand, the hopping terms in the new variables go beyond 
the nearest-neighbor hopping. However, $\lambda_{ll'}$  decrease rapidly with distance,\cite{Chernyshev_93}   
$\lambda _{ll'}\sim 1/|{\bf r}_l - {\bf r}_{l'}|^3$, so the terms beyond the nearest neighbor in the hopping part of 
(\ref{H:symmetric_oxygen_Cu}) and (\ref{H:symmetric_oxygen_Zn}) can be neglected.

The most important result of the transformation 
$\big\lbrace p_{m,\alpha}^x ,\; p_{m\alpha}^y \big\rbrace\rightarrow q_{l,\alpha}(q_{\ell,\alpha})$, is that it allows
to take into account the effects of the the intra-cluster hoppings separately from the local, intra-cluster Cu-O (Zn-O) 
hybridization (Wannier amplitude $\lambda_0$), considered next.

%------------------------------------------------------------------
\subsection{Effective CuO$_{4}$ and ZnO$_4$ states and hoppings}
%------------------------------------------------------------------

Here we list a complete set of hybridized orthogonal electronic states in CuO$_4$ and ZnO$_4$ clusters 
for the one- and two-hole states. We omit the site-index as the states are in the same cluster. 
The states involving antisymmetric oxygen orbitals (\ref{eq:qs}) are excluded since they do not contribute to the 
hybridization. With zero-hole state defined as $|0\rangle$, the CuO$_4$ one-hole states include  
\begin{eqnarray}
|d_{\alpha}\rangle = d_{\alpha}^{\dagger} |0\rangle,\ \ \ |q_{\alpha}\rangle = q_{\alpha}^{\dagger} |0\rangle\, ,
\label{eq:Cu_one_hole_states}
\end{eqnarray}
with energies $\varepsilon_d$ and $\varepsilon_p$, respectively.

The two-hole sector is naturally divided into orthogonal singlet and triplet sectors, to be diagonalized separately,
with  singlet states
\begin{eqnarray}
&&|\psi\rangle = d_\uparrow^\dagger d_\downarrow^\dagger|0\rangle,\ \ \
|\varphi\rangle = q_\uparrow^\dagger q_\downarrow^\dagger|0\rangle,\nonumber\\
&&|\chi\rangle =\frac{1}{\sqrt{2}}
\big(d_\uparrow^\dagger q_\downarrow^\dagger -d_\downarrow^\dagger q_\uparrow^\dagger\big)|0\rangle ,
\label{eq:Cu_two_hole_singlets}
\end{eqnarray}
having energies $2\varepsilon_d+U_d$, $2\varepsilon_p$, and $\varepsilon_d+\varepsilon_p$, respectively, 
and triplets
\begin{eqnarray}
&&|\tau_1\rangle = d_\uparrow^\dagger q_\uparrow^\dagger |0\rangle, \ \ \ 
|\tau_{-1}\rangle = d_\downarrow^\dagger q_\downarrow^\dagger |0\rangle,\nonumber\\
&&|\tau_0\rangle = \frac{1}{\sqrt{2}}\big(d_\uparrow^\dagger q_\downarrow^\dagger 
+ d_\downarrow^\dagger q_\uparrow^\dagger\big)|0\rangle, 
\label{eq:Cu_two_hole_triplets}
\end{eqnarray}
all with the energy $\varepsilon_d+\varepsilon_p$.

Similarly, for ZnO$_4$ cluster the one-hole states are  
\begin{eqnarray}
|a_{\alpha}\rangle = a_{\alpha}^{\dagger} |0\rangle,\ \ \ |q_{\alpha}\rangle = q_{\alpha}^{\dagger} |0\rangle\, ,
\label{eq:Zn_one_hole_states}
\end{eqnarray}
with energies $\varepsilon_{\rm Zn}$ and $\varepsilon_p$, respectively.
The two-hole singlet and triplet states are
\begin{eqnarray}
&&|\psi^{\rm Zn}\rangle = a_\uparrow^\dagger a_\downarrow^\dagger|0\rangle,\ \ \
|\varphi^{\rm Zn}\rangle = q_\uparrow^\dagger q_\downarrow^\dagger|0\rangle, \nonumber\\
&&|\chi^{\rm Zn}\rangle =\frac{1}{\sqrt{2}}\big(a_\uparrow^\dagger q_\downarrow^\dagger -a_\downarrow^\dagger q_\uparrow^\dagger\big)|0\rangle,
\label{eq:Zn_two_hole_singlets}
\end{eqnarray}
with energies $2\varepsilon_{\rm Zn}+U_{\rm Zn}$, $2\varepsilon_p$, $\varepsilon_{\rm Zn}+\varepsilon_p$,
and 
\begin{eqnarray}
&&|\tau_1^{\rm Zn}\rangle = a_\uparrow^\dagger q_\uparrow^\dagger |0\rangle ,\ \ \
|\tau_{-1}^{\rm Zn}\rangle = a_\downarrow^\dagger q_\downarrow^\dagger |0\rangle, \nonumber\\
&&|\tau_0^{\rm Zn}\rangle = \frac{1}{\sqrt{2}}\big(a_\uparrow^\dagger q_\downarrow^\dagger + a_\downarrow^\dagger q_\uparrow^\dagger\big)|0\rangle, 
\label{eq:Zn_two_hole_triplets}
\end{eqnarray}
with energy $\varepsilon_{\rm Zn}+\varepsilon_p$.
The states involving only oxygen orbitals are the same for both ZnO$_4$ and CuO$_4$ clusters. 

Hybridization
terms in ${\cal H}^{\rm loc}$ in (\ref{H:symmetric_oxygen_Cu}) and in ${\cal H}^{\rm loc}_{\rm Zn}$ 
in (\ref{H:symmetric_oxygen_Zn}) originate from hoppings between the orbitals within each cluster 
and lead to mixing of the states in each of the orthogonal sectors of states in 
(\ref{eq:Cu_one_hole_states})-(\ref{eq:Cu_two_hole_triplets}) and 
(\ref{eq:Zn_one_hole_states})-(\ref{eq:Zn_two_hole_triplets}).
Thus, the diagonalization of the local Hilbert space in each  cluster amounts to bringing a few 
$2\!\times\!2$ and $3\!\times\!3$ matrices to a diagonal form. We do not list explicit expressions
for the resulting eigenenergies and eigenvectors, which can be found in Ref.~\onlinecite{Chernyshev_94_01},
and simply assume that they can be easily determined for a given choice of the three-band model parameters.

Since we are interested in the lowest states from each of the $n$-hole sector, we denote them as follows.
The lowest one-hole state of the Cu-cluster is  $|f^{(1)}_{\alpha}\rangle$, which is a normalized linear combination of 
$|d_\alpha\rangle$ and $|q_\alpha\rangle$ in (\ref{eq:Cu_one_hole_states}), with the energy $E_1$. 
The lowest one-hole state for Zn-cluster is  
$|\tilde{f}^{(1)}_{\alpha}\rangle$, a  combination of $|a_\alpha\rangle$ and $|q_\alpha\rangle$ from 
 (\ref{eq:Zn_one_hole_states}), with the energy $\tilde{E}_1$. 
The lowest two-hole states are $|f^{(2)}\rangle$ for Cu-cluster, which is a combination of the singlets 
$|\psi\rangle$, $|\varphi\rangle$, and $|\chi\rangle$ in (\ref{eq:Cu_two_hole_singlets}), with the energy $E_2$ 
and $|\tilde{f}^{(2)}\rangle$ for Zn-cluster, a mix of the singlets in (\ref{eq:Zn_two_hole_singlets}), with the energy $\tilde{E}_2$.

Setting the energy $E_1$ of $|f^{(1)}_{\alpha}\rangle$-state on Cu-cluster to zero, as in Figs.~\ref{fig:energy_level} and
\ref{fig:eff_epsilon}, defines the relevant energy scales of the effective $t$-$\varepsilon$-$U$ model as
$\varepsilon_{\rm eff}^{\rm Zn} = \tilde{E}_1 - E_1$ for the lowest one-hole state of Zn-cluster,
$U_{\rm eff}^{\rm Cu} = E_2 - 2E_1$ as the effective Hubbard gap on Cu-cluster, and 
$U_{\rm eff}^{\rm Zn} = \tilde{E}_2 - 2\tilde{E}_1$ for the effective repulsion on Zn-cluster.
Ignoring the hoppings that involve higher-energy states, the effective $t$-$\varepsilon$-$U$ model, which is
an abbreviated version of the model in (\ref{H:diagonalized_Cu}) and (\ref{H:diagonalized_Zn}), is given by
\begin{eqnarray}
&&{\cal H} = \sum_{l} U_{\rm eff}^{\rm Cu} |f^{(2)}_l\rangle \langle f^{(2)}_l|
\label{H:relevant_Cu}\\
&&\phantom{{\cal H}}+\sum_{\langle ll'\rangle}
\Big\lbrace F^{01}_{10} \left( |f^{(1)}_{\alpha l}\rangle|0_{l'}\rangle\langle f^{(1)}_{\alpha l'}|\langle 0_l | 
+ {\rm H.c.}\right)\nonumber\\
&&\phantom{{\cal H}}+ F^{20}_{11} \left( |f^{(1)}_{\alpha l}\rangle|f^{(1)}_{\alpha l'}\rangle\langle 0_{l'}|\langle f^{(2)}_{l}| 
+ {\rm H.c.}+\left\{l\leftrightarrow l'\right\}\right) \Big\rbrace,\nonumber
\end{eqnarray}
 and
\begin{eqnarray}
&&\delta {\cal H} = \sum_{\ell}\Big\lbrace \big(U_{\rm eff}^{\rm Zn}
+2\varepsilon_{\rm eff}^{\rm Zn}\big) |\tilde{f}^{(2)}_\ell\rangle \langle \tilde{f}^{(2)}_\ell |
+\sum_\alpha\varepsilon_{\rm eff}^{\rm Zn}|\tilde{f}^{(1)}_{\alpha\ell}\rangle \langle \tilde{f}^{(1)}_{\alpha\ell}| \Big\rbrace 
\nonumber\\
&&\phantom{\delta {\cal H}}+\sum_{\langle \ell l\rangle}\Big\lbrace \tilde{F}^{01}_{10} 
\left( |\tilde{f}^{(1)}_{\alpha \ell}\rangle |0_{l}\rangle\langle f^{(1)}_{\alpha l} |\langle 0_{\ell} | + {\rm H.c.}\right)
\label{H:relevant_Zn}\\
&&\phantom{\delta {\cal H}+\sum_{\langle \ell l\rangle}}
+ \tilde{F}^{20}_{11} \left( |\tilde{f}^{(1)}_{\alpha \ell}\rangle|f^{(1)}_{\alpha l}\rangle\langle 0_{l}|\langle \tilde{f}^{(2)}_{\ell} | 
+ {\rm H.c.}\right) \nonumber\\
&&\phantom{\delta {\cal H}+\sum_{\langle \ell l\rangle}}+ \tilde{F}^{02}_{11} 
\left( |\tilde{f}^{(1)}_{\alpha \ell}\rangle|f^{(1)}_{\alpha l}\rangle\langle f^{(2)}_{l}|\langle 0_{\ell} | 
+ {\rm H.c.}\right) \Big\rbrace,\nonumber
\end{eqnarray} 
where the explicit expressions for hopping integrals can be obtained by evaluating matrix elements of the 
Hamiltonian in terms of the original Cu, Zn, and O operators, 
(\ref{H:symmetric_oxygen_Cu}) and (\ref{H:symmetric_oxygen_Zn}),  between the initial and final states in the basis 
of the local eigenstates. For example,
\begin{eqnarray}
F^{01}_{10} = \langle f^{(1)}_{\alpha l}|\langle 0_{l'} | \Big( {\cal H}+\delta {\cal H} \Big) 
|f^{(1)}_{\alpha l'}\rangle|0_{l}\rangle\, .
\end{eqnarray}
In the effective hopping terms shown in Fig.~\ref{fig:effHop} and used in Sec.~\ref{Sec_II} we have employed the
shorthand notations:  $t_1 = F^{01}_{10}$, $t_2 = F^{20}_{11}$, $t_{11} = \tilde{F}^{01}_{10}$, 
$t_{21} = \tilde{F}^{02}_{11}$, and $t_{12} = \tilde{F}^{20}_{11}$.

%------------------------------------------------------------------
\section{Details of the $T$-matrix calculation}
\label{AppB}
%------------------------------------------------------------------

Here we provide some of the technical details of the $T$-matrix approach, which 
follows closely Ref.~\onlinecite{Chernyshev_02}. 

The $2\times 2$ matrix Green's function of a magnon in the square-lattice antiferromagnet can be written 
in a Dyson-Belyaev form:
\begin{eqnarray}
&&\hat{G}_{\bf k}(\omega) = \frac{-1}
{\big(\omega-\omega_{\bf k}-\Sigma_{\bf k}^{11}\big)
\big(\omega+\omega_{\bf k}+\Sigma_{\bf k}^{22}\big)
+\big(\Sigma_{\bf k}^{12}\big)^2}\nonumber\\
&&\phantom{\hat{G}_{\bf k} (\omega)=}\times
\begin{pmatrix} 
-\omega -\omega_{\bf k}-\Sigma_{\bf k}^{22} &\Sigma_{\bf k}^{12}\\ 
\Sigma_{\bf k}^{21}&\omega-\omega_{\bf k}-\Sigma_{\bf k}^{11}
\end{pmatrix}  .
\label{eq:Greens_function}
\end{eqnarray}
The low-doping consideration requires an expansion of (\ref{eq:Greens_function}) in powers of $\Sigma$,
which given by 
\begin{eqnarray}
&&G^{11} = G^{0,11} + G^{0,11}\Sigma^{11}G^{11} + G^{0,11}\Sigma^{12}G^{21}\nonumber\\
 &&\phantom{G^{11}}\simeq G^{0,11} + G^{0,11}
 \Sigma^{11}G^{0,11} + \mathcal{O}(x^2)\, , \label{eq:Greens_function_elements}\\
&&G^{12} = G^{0,11}\Sigma^{11}G^{12} + G^{0,11}\Sigma^{12}G^{22}\nonumber\\
&&\phantom{G^{12}}\simeq G^{0,11}
\Sigma^{12}G^{0,22} + \mathcal{O}(x^2)\nonumber
\end{eqnarray}
where we drop the common ${\bf k}$ and $\omega$ dependencies for shorthand notations and the 
non-interacting magnon  Green's function is
\begin{eqnarray}
\hat{G}_{\bf k}^0(\omega) = 
\begin{pmatrix} \displaystyle
\frac{1}{\omega-\omega_{\bf k}+{\it i}0} &  \displaystyle 0 \\ 
 \displaystyle 0 &  \displaystyle \frac{-1}{\omega + \omega_{\bf k}-{\it i}0}
\end{pmatrix} ,
\end{eqnarray}
where we normalize all the energies to $\Omega_0 = 4SJ$  and the magnon frequency is 
$\omega_{\bf k} = \sqrt{1-\gamma_{\bf k}^2}$, with $\gamma_{\bf k}=\left( \cos{k_x} + \cos{k_y} \right)/2$.

The self-energies in (\ref{eq:Greens_function}) and (\ref{eq:Greens_function_elements}) 
are linear in the doping concentration $x$ and are related to the 
forward-scattering components of the $T$-matrix via
\begin{eqnarray}
{\Sigma}^{ij}_{\bf k}(\omega) = 
x\sum_{\mu , I}\delta_{{\bf k},{\bf k}'} {{T}}^{I\mu,ij}_{{\bf k},{\bf k}'}(\omega)\, ,
\label{eq:self_energy_appB}
\end{eqnarray}
where the sum is over the sublattice index  ($I = A$, $B$) 
and $\mu = s$-, $p$-, and $d$-wave components.

Using the sublattice-$A$ as an example,\cite{Chernyshev_02} the $s$-,  $p$-, and $d$-wave components of the 
$T$-matrix are
\begin{eqnarray}
\hat{T}_{{\bf k},{\bf k}'}^{A\mu}(\omega) &=& \hat{\mathcal{V}}_{{\bf k},{\bf k}'}^{A\mu} 
\Gamma_{\mu}(\omega)\;\;\;\;(\mu \in p,d)\, ,
\label{eq:Tmatrices}\\
\hat{T}^{As}_{{\bf k},{\bf k}'}(\omega) &=& \hat{\mathcal{V}}^{As}_{{\bf k},{\bf k}'}\Gamma_s(\omega) 
-\omega|\Delta s_{\bf k}^A\rangle\otimes\langle\Delta s_{{\bf k}'}^A| \nonumber\\
&&+|s_{\bf k}^A\rangle \otimes\langle\Delta s_{{\bf k}'}^A| 
+ |\Delta s_{\bf k}^A\rangle\otimes\langle s_{{\bf k}'}^A|\, ,\nonumber
\end{eqnarray}
where we use $\otimes$ to denote the direct product of the column and row vectors and 
$\hat{\mathcal{V}}_{{\bf k},{\bf k}'}^{A\mu}$ stand for 
the $s$-, $p$-, and $d$-wave components of the scattering potential that can be written as 
\begin{eqnarray}
\hat{\mathcal{V}}^A_{{\bf k},{\bf k}'} = \sum_{\mu\in s,p,d} C^\mu|\mu_{\bf k}\rangle\otimes\langle\mu_{{\bf k}'}| \, ,
\label{eq:Vs}
\end{eqnarray}
in which the $s$-, $p$-, and  $d$-wave vectors are given by
\begin{eqnarray}
&&\langle s_{\bf k}^A| = \omega_{\bf k} \begin{pmatrix} u_{\bf k}, &\ -v_{\bf k} \end{pmatrix} ,\;\;\;\;\;\;\;
\langle\Delta s_{\bf k}^A| = \begin{pmatrix} u_{\bf k}, & \ v_{\bf k} \end{pmatrix},
\label{eq:wave_vectors_A}\\
&&\langle p_{k_{x(y)}}^A| = \frac{1}{\sqrt{2}}\sin{k_{x(y)}} \begin{pmatrix} v_{\bf k}, & \ u_{\bf k}\end{pmatrix},\;\;\;
\langle d_{\bf k}^A| = \gamma_{\bf k}^- \begin{pmatrix} v_{\bf k}, &\ u_{\bf k} \end{pmatrix},\nonumber
\end{eqnarray} 
with the Bogolyubov parameters $u_{\bf k} = \sqrt{(1+\omega_{\bf k})/2\omega_{\bf k}}$ and 
$v_{\bf k} = -{\rm sgn}(\gamma_{\bf k})\sqrt{(1-\omega_{\bf k})/2\omega_{\bf k}}$. The $\omega$-dependent
resolvents $\Gamma_\mu(\omega)$  in (\ref{eq:Tmatrices}) are listed below, see (\ref{eq:gamma_functions}).

The coefficients in (\ref{eq:Vs}) contain the dependence on the frustrating terms in the dilution-frustration model
\begin{eqnarray}
&&C^{s} = 1\, ,\nonumber\;\;\;\;\;
C^{p} = 1+ 2\frac{J_{\rm Zn}'}{J}+2\frac{J_{\rm Zn}''}{J}\, ,\;\;\;\;\;
C^{d} = 1+4\frac{J_{\rm Zn}'}{J}\, .
\label{eq:Cs}
\end{eqnarray}
For the sublattice-$B$, the equivalent expressions are obtained via 
$\hat{T}^{B\mu}_{{\bf k},{\bf k}^\prime}(\omega)= 
\hat{T}^{A\mu}_{{\bf k},{\bf k}^\prime} (-\omega)\{u\leftrightarrow v\}$, see Ref.~\onlinecite{Chernyshev_02}.

The resolvents $\Gamma_\mu(\omega)$ are
\begin{eqnarray}
&&\Gamma_s (\omega)= \frac{\big(1+\omega\big)\rho(\omega)}{1-\omega\big(1+\omega\big)\rho(\omega)}\, ,
\nonumber\\
&&\Gamma_p (\omega)= \frac{-2}{2-C^p\big(1-\omega\big)
\big[ 1 -\omega^2\rho(\omega) + \rho_d(\omega)\big]}\, , \label{eq:gamma_functions}\\
&&\Gamma_d (\omega)= \frac{-1}{1+C^d\big(1-\omega\big)\rho_d(\omega)}\, ,\nonumber
\end{eqnarray}
where the functions $\rho(\omega)$ and $\rho_d(\omega)$  are
\begin{eqnarray}
&&\rho(\omega) = \sum_{\bf k} \frac{1}{\omega^2 - \omega_{\bf k}^2},\;\;\;\;
\rho_d(\omega) = \sum_{\bf k} \frac{\big( \gamma_{\bf k}^-\big)^2}{\omega^2 - \omega_{\bf k}^2},
\label{eq:rhos}
\end{eqnarray}
with   $\gamma_{\bf k}^- = ( \cos{k_x} - \cos{k_y})/2$. 
They can be expressed through the complete
elliptic integrals of the first and second kind.\cite{Chernyshev_02}

 The self-energy matrix elements 
are related via $\Sigma_{\bf k}^{22}(\omega) = \Sigma_{\bf k}^{11}(-\omega)$ 
and $\Sigma_{\bf k}^{12}(-\omega) = \Sigma_{\bf k}^{21}(\omega)$.
 The partial-wave terms in the self-energies are 
\begin{eqnarray}
\frac{\hat{\Sigma}^s_{{\bf k}}(\omega)}{xC^s\omega_{\bf k}} &=&  
\Gamma_s^+(\omega)
\begin{pmatrix} 1&\gamma_{\bf k}\\ \gamma_{\bf k} &1\end{pmatrix} 
 -\frac{\omega}{\omega_{\bf k}}
 \begin{pmatrix}1&0\\0&-1\end{pmatrix}
 \nonumber\\
&+&\Gamma_s^-(\omega)
\begin{pmatrix} \omega_{\bf k} &0 \\ 0 & -\omega_{\bf k} \end{pmatrix}
+2\begin{pmatrix} 1 & 0\\ 0 &1 \end{pmatrix}\, ,
\end{eqnarray}
for the $s$-wave,
\begin{eqnarray}
\frac{\hat{\Sigma}^p_{{\bf k}}(\omega)}{xC^p\omega_{\bf k}} &=& 
\bigg[1-\bigg(\frac{\gamma_{\bf k}^-}{\omega_{\bf k}}\bigg)^2\bigg]
\bigg\lbrace\Gamma_p^+(\omega)
\begin{pmatrix} 1&-\gamma_{\bf k}\\-\gamma_{\bf k} & 1\end{pmatrix}\nonumber\\
&&\phantom{\bigg[1-\bigg(\frac{\gamma_{\bf k}^-}{\omega_{\bf k}}\bigg)^2\bigg]}
+\Gamma_p^-(\omega)\begin{pmatrix}-\omega_{\bf k} &0\\0&\omega_{\bf k}\end{pmatrix}\bigg\rbrace
\, ,\ \ \ \ \ 
\end{eqnarray}
for the $p$-wave, and 
\begin{eqnarray}
\frac{\hat{\Sigma}^d_{{\bf k}}(\omega)}{xC^d\omega_{\bf k}} &=& 
\bigg(\frac{\gamma_{\bf k}^-}{\omega_{\bf k}}\bigg)^2
\bigg\lbrace
\Gamma_d^+(\omega)\begin{pmatrix} 1&-\gamma_{\bf k}\\-\gamma_{\bf k} & 1\end{pmatrix}\nonumber\\
&&\phantom{\bigg(\frac{\gamma_{\bf k}^-}{\omega_{\bf k}}\bigg)^2} 
+\Gamma_d^-(\omega)
\begin{pmatrix}-\omega_{\bf k} &0\\0&\omega_{\bf k}\end{pmatrix}
\bigg\rbrace\, ,
\end{eqnarray}
for the $d$-wave.
Here $\Gamma_\mu^{\pm} = \frac{1}{2}\left[\Gamma_\mu(\omega)\pm\Gamma_\mu(-\omega)\right]$ and 
$\hat{\Sigma}^p_{{\bf k}}(\omega) = \hat{\Sigma}^{p_x}_{{\bf k}}(\omega) 
+ \hat{\Sigma}^{p_y}_{{\bf k}}(\omega)$.

%------------------------------------------------------------------------------------------

\end{document}